\documentclass[a4paper,11pt]{article}
\pdfoutput=1 % if your are submitting a pdflatex (i.e. if you have images in pdf, png or jpg format)
\usepackage{jcappub}
\usepackage[T1]{fontenc}
\usepackage{amssymb,amsmath,amsfonts}
\usepackage{hyperref}
\usepackage{float}
\usepackage{graphicx}
\usepackage{xcolor}
\usepackage{tensor}
\usepackage[utf8]{inputenc}
\usepackage{caption,subfig}
\usepackage{feynmf}
\usepackage{tikz}
\usepackage{nicefrac}

\usepackage{dcolumn}% Align table columns on decimal point
\usepackage{bm}% bold math
\usepackage{esvect}
\usepackage{accents}

\usepackage{hyperref}
\hypersetup{colorlinks=true,linkcolor=magenta,anchorcolor=green,citecolor=cyan,filecolor=black,menucolor=black,urlcolor=brown}
\usepackage{slashed}
\newcommand{\be}{\begin{equation}}
\newcommand{\ee}{\end{equation}}
\newcommand{\ba}{\begin{eqnarray}}
\newcommand{\ea}{\end{eqnarray}}

\def\mpl{M_{\rm Pl}}

\title{Unveiling the Galileon in a three-body system : scalar and gravitational wave production }
\author[a]{Philippe Brax,}
\author[b]{Lavinia Heisenberg,}
\author[c]{Adrien Kuntz}
\affiliation[a]{Institut de Physique Th\'eorique, Universit\'e  Paris-Saclay, CEA, CNRS, F-91191 Gif-sur-Yvette Cedex, France}
\affiliation[b]{Institute for Theoretical Physics,
	ETH Zurich, Wolfgang-Pauli-Strasse 27, 8093, Zurich, Switzerland}
\affiliation[c]{Aix Marseille Univ, Universit\'{e} de Toulon, CNRS, CPT, Marseille, France}
\emailAdd{philippe.brax@ipht.fr}
\emailAdd{lavinia.heisenberg@phys.ethz.ch}
\emailAdd{kuntz@cpt.univ-mrs.fr}

\abstract{

We consider the prospect of detecting  cubic Galileons through their imprint on  gravitational wave signals from a triple system. Namely, we consider a massive Black Hole (BH) surrounded by a binary system of two smaller BHs. We assume that the three BHs acquire a conformal coupling to the scalar field whose origin could be due to cosmology or to the galactic environment. In this case, the massive BH has a Vainshtein radius which englobes the smaller ones and suppresses the scalar effects on the motion of the binary system. On the other hand the two binaries can be outside each other's redressed Vainshtein radius calculated in the background of the central BH, allowing for a perturbative treatment of their dynamics. Despite the strong Vainshtein suppression, we find that the scalar effects on the binary system are slightly enhanced with respect to the static case and a significant amount of power can be emitted in  the form of the Galileon scalar field, hence actively participating in the inspiralling phase. We compute the modification to the GW phase and show that it can lead to a detectable signal for large enough effective scalar coupling.

}

\begin{document}
	\allowdisplaybreaks[1]
	\maketitle
	\flushbottom
	%
	%------------------------------------------------------------------------------

\section{Introduction}
In 1915 Albert Einstein formulated his theory of General Relativity (GR), which still remains  our best description of gravitational phenomena. Einstein ingeniously postulated the existence of a geometrical formulation of gravity inferred from  the equivalence principle and chose to describe gravitational effects as induced by the spacetime curvature (even though alternative geometrisations exist \cite{BeltranJimenez:2019tjy}). In this formulation, mass and energy causes  spacetime to be curved and the dynamics are described by Einstein's field equations. Its non-relativistic limit naturally coincides with Newton's gravity. Since its inception GR has withstood  intense scrutiny and has been experimentally confirmed on a multitude of scales. Three fundamental effects were suggested by Einstein as a way of testing GR: the gravitational deflection of light, the gravitational redshift and the perihelion shift. They were all successfully confirmed. Another important prediction of Einstein's theory is the existence of Black Holes (BHs). Objects like neutron stars or white dwarfs are the standard outcome of stellar evolution but under unusual circumstances with star masses exceeding a certain upper limit of order   20 solar masses BHs can form. A breathtaking recent event was the first picture of a BH as released by the Event Horizon Telescope \cite{Akiyama:2019cqa}. One defining property of BHs within the realm of GR is the no-hair theorem, which states that the created BH after collapse  depends only on its mass, its electric charge and its angular momentum.

The successes of GR do not stop here. A fundamental prediction of GR is the existence of gravitational waves (GWs) in the presence of a time-varying gravitational quadrupole moment. A linear analysis is enough to reveal  the main features of their evolution and propagation . For sufficiently weak gravitational fields, the metric manifests itself as a small perturbation and the linearised Einstein's field equations are wave equations. The evolution of the GWs is then obtained in terms of the retarded Green's function and  transverse waves that propagate at the speed of light. Realistic GWs  follow highly non-linear evolutions beyond the linear regime and in some cases even form intrinsic spacetime singularities. The radiative sector of GR was hidden to observations until recently. Only indirect evidence for the presence of GWs followed from  the measurements on the Hulse-Taylor double pulsar as a manifestation of period variation \cite{Wex:2014nva}. The remarkable breakthrough of the first direct detection of GWs as reported in 2016 by the LIGO team \cite{Abbott:2016blz} opened up a new window for astrophysical observations, combined with a wealth of exciting, otherwise barely accessible phenomena. Quite generically, one can source three different type of GWs: shock waves, periodic waves and stochastic waves. While rotating stars, such as binary pulsars, generate periodic GWs, the massive collision of BHs and neutron stars causes shock waves. A new era of multimessenger data arose with the first detection of the merger of two neutron stars, where both the GWs as well as the electromagnetic signal were observed at the same time \cite{TheLIGOScientific:2017qsa}. Amongst many outstanding scientific implications, one immediate consequence was an incredibly tight constraint $\Delta c=10^{-15}$ on the difference in the propagation speeds of GWs and photons. As a result, dark energy models featuring an anomalous propagation speed for GWs are highly disfavoured by this gravitational observation \cite{Brax:2015dma,Lombriser:2016yzn,Ezquiaga:2017ekz,Creminelli:2017sry,Sakstein:2017xjx,Baker:2017hug} (recent reviews \cite{Heisenberg:2018vsk,Ezquiaga:2018btd,Kase:2018aps}).

Even though GR describes in an exemplary manner most of the gravitational  phenomena which are accessible to observations, it faces some tenacious challenges. As a non-renormalisable theory its UV completion  into a quantum gravity is unknown.  Furthermore, curvature singularities in form of Big Bang and BH singularities can emerge. Cosmological observations enforce the necessity for unknown matter and energy forms, like the inflaton field, dark matter and dark energy. Another persistent challenge is the cosmological constant problem. Some of these problems are addressed by extensions of GR. Prominent classes of such extensions are scalar-tensor \cite{Horndeski:1974wa}, vector-tensor \cite{Heisenberg:2014rta,Jimenez:2016isa} and tensor-tensor \cite{deRham:2010kj} theories. A common property that all these theories share is the presence of Galileon interactions for the helicity-0 mode in some limits. The scalar Galileon model \cite{Nicolis:2008in} is very special in the sense that  derivative self-interactions result in equations of motion that are at most second order. Furthermore, they are technically natural, meaning that the classical coefficients of the interactions receive no quantum corrections \cite{Nicolis:2004qq,Hinterbichler:2010xn,dePaulaNetto:2012hm,Heisenberg:2014raa,Heisenberg:2019udf,Heisenberg:2019wjv}. The presence of derivative self-interactions equip such theories with the  Vainshtein screening mechanism. As a result, in the vicinity of matter, the non-linear interactions of the Galileon scalar become large and hence suppress its coupling to matter \cite{Vainshtein:1972sx}. Since the discovery  of the Galileon interactions there has been a flurry of works  related to Galileon cosmology \cite{Chow:2009fm,DeFelice:2010nf,deRham:2011by,deRham:2010tw}, inflation \cite{Creminelli:2010ba,LevasseurPerreault:2011mw,Kobayashi:2010cm,Burrage:2010cu}, laboratory tests \cite{Brax:2011sv}, BHs \cite{Babichev:2016fbg,Babichev:2015rva}, lensing \cite{Wyman:2011mp}, superluminal propagation around compact sources \cite{Goon:2010xh,deFromont:2013iwa}.

The phenomenological viability of these models should be tested  on small and large scales. Given the successful developments in GW's astronomy in recent years, it has become possible to look for such direct signals. Besides the existing observations of LIGO and Virgo, the planned LISA project \cite{Seoane:2013qna} aims at pushing further  our current observational boundaries. LISA will be an interferometric GW detector in space. Three satellites will form a triangle with a side length of 2.5 million kilometers in order to search for GWs with laser interferometers. LISA will be most sensitive in the frequency range between $f\sim3 \times 10^{-5}$ and  $f\sim10^{-1}$ Hz with a relative precision in frequency of order $\delta f /f \sim 10^{-8}$. In this sense, it differs from detectors installed on earth, which can only examine higher frequencies. LISA will be sensitive to GWs from super-heavy BHs in a large part of the observable universe and will therefore be much more sensitive than earth detectors like LIGO. It may also be possible to detect waves that originate from the Big Bang.

As a  phenomenological test of the existence of scalar fields, BHs usually do not present many features and are deemed to be  irrelevant probes because of the no-hair theorem \cite{Hawking1972,Hui_2013}. However there are many ways out of the no-hair theorem. Indeed the  time-dependence of a scalar field at spatial infinity induces scalar hair around BHs \cite{Jacobson:1999vr}. This time-dependence could be due to cosmological boundary conditions or to the environment in which the BHs are located \cite{Horbatsch:2011ye}. Concerning the Galileon, we will review in Section \ref{sec:BH_hair} how hair can be  induced by an asymptotic timelike gradient of the scalar, showing that it can even lead to the generation of large scalar couplings.  For simplicity only the cubic interaction will be considered here but this will already capture all the typical features of the Galileons.

Even if one circumvents the no-hair theorem, the modification of the GW signal of inspiralling binaries by the Galileon field is usually assumed to be too small to be observable because the field is screened by the Vainshtein mechanism \cite{deRham:2012fg,Dar:2018dra,deRham:2012fw}. Despite the fact that scalar-tensor theories generically predict dipole radiation which is enhanced with respect to the quadrupole in the post-Newtonian expansion \cite{Eardley1975ApJ}, the Vainshtein suppression is so enormous that any deviation from GR would be tiny. However, GW detectors such as LISA can monitor the inspiral of compact objects during a large number of GW cycles, thereby greatly enhancing the potentiality of detecting any deviation from GR. It is thus worth investigating  the detectability of a Galileon field by LISA.

 An immediate difficulty to be  faced when studying Galileon dynamics is that the field equations are highly nonlinear (from the very existence of the Vainshtein mechanism) and post-Newtonian methods such as in standard scalar-tensor theories \cite{Kuntz:2019zef} are inappropriate. However, in the case of an extreme mass ratio inspiral (EMRI) of a solar-size BH interacting with a supermassive one, one can take advantage of the extreme mass ratio to set up a perturbative calculation. In this article, we will even go one step further and consider a triple system constituted of two 'small' BHs in a binary system orbiting a giant BH at the centre of a galaxy. We will be able to solve perturbatively for the global motion of the binary around the central BH, as well as for the motion of the binary system itself around its centre-of-mass.  This is due to the fact that, while being fully inside the Vainshtein radius of the large BH, the binary system can lie outside its own 'redressed' Vainshtein radius\footnote{In the background of the central BH, the Vainshtein radius of the two BHs is modified as the effective conformal coupling, after canonically normalising the scalar field, is reduced. The redressing factor depends on the background field generated by the central BH.}. This property allows us to set up a perturbative calculation in which  the quadratic action in the background field of the massive BH dominates over the cubic interactions.

Before giving more  details of our calculations, let us present a few order-of-magnitude estimates characteristic of  our system. We will consider a massive BH of mass $m_0 \sim 10^6 M_\odot$ at the centre of a galaxy, and two 'small' BHs of mass $m_1, m_2 \sim 30 M_\odot$ orbiting in its vicinity. For simplicity, all trajectories are assumed to be circular, i.e the two BHs are in a circular orbit of radius $d$ and their common barycentre is in circular orbit of radius $r$ around the massive BH, with $d \lesssim r$. A more realistic treatment would necessitate to take  into account the eccentricity of the orbits which can grow to significant values in this kind of configurations \cite{Randall:2019sab,Randall:2017jop,Randall:2018nud}. At a frequency of approximately $\omega = 10^{-3}$ Hz and higher, the two BHs could emit gravitational waves potentially detectable in the LISA band provided that their amplitude is large enough. This frequency corresponds to a maximal separation of the binary system  of  $d_\mathrm{max} = (Gm_1/\omega^2)^{1/3} \sim 10^9$ m $\sim 0.01$ AU, using Kepler's third law. Higher frequencies lead to shorter distances. On the other hand, we will see in Section \ref{sec:setup} that in order for our perturbative calculation to be valid,  one should require
\begin{equation}
\left(\frac{\beta_\mathrm{binary} m_1}{ \beta_0 m_0} \right)^{1/3} r \lesssim d \lesssim \left(\frac{m_1}{m_0} \right)^{1/3} r  \; ,
\end{equation}
so that the binary system should be further than $0.5$ AU from the central black hole. Here $\beta_0$ is the coupling of the scalar to the central BH and $\beta_\mathrm{binary}$ the couplings to the smaller ones. The corresponding Vainshtein suppression of the fifth force can be as low as  $(r/r_*)^{3/2} \sim 10^{-16}$ where $r_*$ is the Vainshtein radius of the central BH. Despite this strong suppression, we will show that the Galileon field can lead to observable changes in the GW phase provided one chooses moderately large values for the effective scalar coupling.
 We then compute both the two-body energy and the power dissipated from the system. While the former involves the usual Vainshtein suppression and leads to negligible departures from GR, the latter contains non-trivial powers of the lengthscales of the system which makes it non-negligible in this particular configuration. We finally derive the modification of the binary dynamics when the scalar power is subdominant, and show that it could be observable thanks to the large number of gravitational wave cycles.

 Our article is structured as follows. In Section \ref{sec:setup} we present both the theory and the physical system which we consider. We then give the essence of our perturbative calculation. In Section \ref{sec:Green} we explicitly compute the Green's function of the cubic Galileon. This will be used in Section \ref{sec:static} to compute the Galileon's correction to the energy, as well as in Section \ref{sec:diss_power} to calculate the power dissipated from the system to lowest order. Finally, we present in Section \ref{sec:inspiral} the correction to the GW phase induced by the scalar dissipated power found in the preceding Section. We use units in which $\hbar = c = 1$, we define Planck's mass by $\mpl^2 = 1/(8 \pi G)$ where $G$ is Newton's constant, and our metric convention is $(-+++)$.

% For larger values of $\beta$, the two BH's of the binary system first emit scalar power  dominantly before the GW emission due to the orbital motion of the binary system around the central BH  becomes dominant. In this regime, the scalar emission simply delays the plunge  of the two BH's towards the central one.
%These scenarios require large couplings which could emerge from the time dynamics of the BH's in the galactic environment. We will not try to give a detailed account
%of the possible values of the conformal couplings, leaving it for future work.

%
%
%\section{Some orders of magnitude} \label{sec:order_of_mag}

%LISA band : $3 \times 10^{-5}$ to $10^{-1}$ Hz. LISA relative precision in the frequency : $\delta f /f \sim 10^{-8}$.

%For $30 M_\odot$ black holes (which are among the targets of LISA, sufficiently massive to be observed at large distances), this translates via Kepler's law into a maximal separation : $d_\mathrm{max} = (GM/\omega^2)^{1/3} \sim 0.1 \mathrm{AU} \sim 10^{10}$ m.

%With the condition $d \gtrsim r/50$ this means that we should look for binary objects closer than $5$ AU of the central supermassive black hole.
%
%Finally, the Vainshtein radius of a solar mass object is $10^3$ Pc $\sim 10^{19}$ m, the one of a supermassive black hole of $10^6 M_\odot$ is $10^5$ Pc $\sim 10^{21}$ m and the associated Vainshtein suppression is $(r/r_*)^{3/2} \sim 10^{-15}$ for a distance of $5$ AU. This is quite smaller than LISA precision...

\section{A three-body system in the cubic Galileon} \label{sec:setup}
In this section we will present the theory setup and the configuration that we will use throughout this work. We will make explicit our chosen notations and units.

\subsection{The theory}
We  assume that the gravitational interactions are successfully described by the laws of GR. We will further consider the presence of an additional massless scalar field $\pi$ with a derivative self-interaction. The total action defining our theory is
\begin{equation} \label{eq:base_action}
S = \frac{1}{2} \int \mathrm{d}^4 x \sqrt{-g} \left[ \mpl^2 R - (\partial \pi)^2 - \frac{1}{\Lambda^3} (\partial \pi)^2 \square \pi \right] + S_m[\tilde g_{\mu \nu}, \psi_i] \; ,
\end{equation}
where $g$ denotes the determinant of the metric, $R$ the Ricci scalar and $\Lambda$ is the energy scale of the cubic Galileon interaction. In order to give a rough estimate for the latter, if the Galileon is supposed to contribute to the accelerated expansion of our universe one should impose $\Lambda^3 \sim H^2 \mpl$. Nevertheless, we will keep this scale arbitrary and assume that a cosmological constant lies behind the acceleration. The metric $\tilde g_{\mu \nu}$ is the Jordan frame metric, which couples to matter. We take it to be simply conformally related to the Einstein frame metric, $\tilde g_{\mu \nu} = A^2(\pi) g_{\mu \nu}$ where $A(\pi) = e^{\beta \pi / \mpl}$.
The contribution of the derivative self-interaction to the scalar field equation of motion is of second order
\begin{equation}
\frac{\delta S}{\delta \pi}\supset [\Pi]+\frac{1}{\Lambda^3}\left( [\Pi]^2-[\Pi^2]\right)-\frac{1}{\Lambda^3}R_{\mu\nu}\partial^\mu\pi\partial^\nu\pi,
\end{equation}
where $[\Pi]$ represents the trace of $\Pi_{\mu\nu}=\nabla_\mu\partial_\nu\pi$. The Einstein's equations are given by
\begin{equation}
\mpl^2 G_{\mu\nu}=T_{\mu\nu}^\pi+ T_{\mu\nu}^m
\end{equation}
where $T_{\mu\nu}^m$ is the matter energy-momentum tensor, and the associated stress energy tensor of the Galileon field is given in (\ref{eq:EMT}).
One can see from the above expressions that the equations of motion remain second order for both the metric and the scalar field despite the presence of derivative self-interactions.

%Note the ghostlike kinetic term in eq. \eqref{eq:base_action} : as shown in Ref. \cite{Babichev_2013},  this is required once we consider asymptotic de Sitter boundary conditions for a pointlike object with scalar charge. We briefly recall the results of this reference in the next subsection.
%

\subsection{BH physics of the Galileon} \label{sec:BH_hair}

BHs in the presence of a cubic Galileon interaction have been studied extensively \cite{Babichev:2016fbg,Babichev:2015rva,Babichev:2013cya,Babichev_2013}. In particular, Ref. \cite{Babichev:2016fbg} showed that hairy solutions do exist once we impose cosmological boundary conditions. The scalar charge of massive objects in a cubic Galileon was studied in \cite{Babichev_2013} where it was shown that, even starting from a negligible bare coupling of the scalar to matter $\beta \sim 0$, an order-one \textit{effective} scalar charge $\beta_\mathrm{eff}$ emerges from  the  cosmological boundary condition. Let us see in more details how this effect arises.

As shown in \cite{Babichev_2013, Babichev:2016fbg} the scalar field equation can be written as a conservation of a current $\nabla_\mu J^\mu = 0$ with
\begin{equation}
J^\mu = \partial^\mu \pi + \frac{1}{\Lambda^3} \square \pi \partial^\mu \pi - \frac{1}{2 \Lambda^3} \nabla^\mu \left( (\partial_\nu \pi)^2  \right) \; .
\end{equation}
We consider a vacuum spherically symmetric solution of the field equations with an ansatz for the scalar field
\begin{equation}
\pi = q t + \bar \pi(r)\; ,
\end{equation}
where $q= m_{\rm Pl} t_{\rm scalar}^{-1}$ is the time derivative of the field, and a static and spherically symmetric ansatz for the metric is chosen
\begin{equation}
ds^2 = - e^{\nu(r)} dt^2 + e^{\lambda(r)} dr^2 + r^2 d\Omega^2 \; .
\end{equation}
Notice that the value of $q$ is arbitrary here. When a BH is embedded in a galactic environment, the time dependence of the scalar could be due to galactic phenomena which take place on shorter time scales than the Hubble rate $t_{\rm scalar} < H_0^{-1}$. In the following, we will take $q$ as a phenomenological parameter.
The $(tr)$ component of the metric equations is then equivalent to $J^r = 0$. We further make the assumption that the scalar is a test  field, i.e. we neglect its backreaction on the metric. We seek for solutions perturbatively close to the Schwarzschild one. With this supplementary assumption, Ref. \cite{Babichev_2013} then showed that the scalar field solution to the $J^r = 0$ equation is
\begin{equation} \label{eq:sol_phiprime_BEF}
\pi' = - \frac{1}{4} \Lambda^3 r \left( 1 - \sqrt{1 + \frac{8 M q^2}{8 \pi \mpl^2 r^3 \Lambda^6}} \right) \left[ 1 + \mathcal{O}\left( \frac{r_s}{r} \right) \right] \; ,
\end{equation}
where $M$ is the ADM mass of the BH, and the solution has been expanded outside the Schwarzschild radius $r_s$.
 We can define an effective scalar charge and its associated Vainshtein radius,
\begin{equation} \label{eq:def_vainshtein_radius}
\beta_\mathrm{eff} = \frac{q^2}{2 \Lambda^3 \mpl}, \quad r_V^3 = \frac{\beta_\mathrm{eff} M}{8 \pi \mpl \Lambda^3} \; ,
\end{equation}
such that for $r \ll r_V$ the solution reads
\begin{align} \label{eq:phi_SS}
\pi' =  \left( \frac{\beta_\mathrm{eff} M \Lambda^3 }{8 \pi \mpl r} \right)^{1/2}
 =  \frac{\beta_\mathrm{eff} M}{8 \pi \mpl r_V^2} \left( \frac{r_V}{r} \right)^{1/2} \;.
\end{align}
This solution is exactly the field generated by a massive body coupled with a Jordan frame metric $A(\pi) = e^{\beta_\mathrm{eff} \pi / \mpl}$. We will consequently model our BHs with a point-particle action with coupling $\beta_\mathrm{eff}$ as
\begin{equation} \label{eq:matter_pp_action}
S_m = - M \int \mathrm{d}t \; e^{\beta_\mathrm{eff} \pi / \mpl} \sqrt{- g_{\mu \nu} v^\mu v^\nu} \; ,
\end{equation}
where $v^\mu = \frac{dx^\mu}{dt}$ is the four-velocity of the BH.

It is important to notice that for cosmological boundary conditions, $q \sim H \mpl$ and for $\Lambda$ related to the dark energy scale, $\Lambda^3 \sim H^2 \mpl$, then the effective scalar charge is close to unity. The associated Vainshtein radius is, for an object of solar mass, of order of a kiloparsec. This leads to a Vainshtein suppression of the fifth force
\begin{equation}
\frac{\pi}{\phi_N} \sim \beta_\mathrm{eff} \left( \frac{r}{r_V} \right)^{3/2}\;,
\end{equation}
where $\phi_N = M / (4 \pi \mpl r)$ is the Newtonian potential.

 However, in this article we will be interested in the case where $q$ does not arise from a cosmological boundary condition but rather from time-dependent phenomena in the environment of the BHs. In traditional scalar-tensor theories, this 'Miracle Hair Growth Formula' relating the temporal variation of the scalar far from the system to the scalar charge was found by Jacobson \cite{Jacobson:1999vr}. It has been used to predict dipole radiation from binary black holes systems \cite{Horbatsch:2011ye}. In our case, the only condition which needs to be satisfied is that the Vainshtein radius associated to this asymptotic gradient, given in eq. \eqref{eq:def_vainshtein_radius}, should be much larger than the typical size of the system we will consider.  When such scalar couplings are $\beta_\mathrm{eff} \gg 1$, this enhances the detectability of the modification of gravity considered here.

%Alternatively, one can also translate the results we will obtain in this article as arising from \textit{any} temporal gradient $q$ of a cubic Galileon scalar field far from the system, be it due to cosmology or to the local environment in which the system is located. In traditional scalar-tensor theories, this 'Miracle Hair Growth Formula' found by Jacobson \cite{Jacobson:1999vr} has been used to predict dipole radiation from binary black holes systems \cite{Horbatsch:2011ye}. The only condition which needs to be satisfied is that the Vainshtein radius associated to this asymptotic gradient, given in eq. \eqref{eq:def_vainshtein_radius}, should be much larger than the typical size of the system we will consider. In this case, one can achieve scalar couplings $\beta_\mathrm{eff} \gg 1$ which would enhance the detectability of the modification of gravity considered here.

\subsection{Redressing the interactions} \label{sec:three_body_problem}

Emission of gravitational and scalar waves by a system of binary pulsars in the presence of Galileon interactions has been studied in Refs. \cite{deRham:2012fg,deRham:2012fw,Dar:2018dra}. In this article, we will consider a different physical situation amenable to a perturbative treatment.
The system we will study is a three-body system, which is composed of a massive BH of mass $\mathcal{O}(10^6) M_\odot$ at the origin of the coordinates, and two 'small' BHs of mass $\mathcal{O}(10) M_\odot$. We will generically denote distances to the massive BH  by $r$, and distances between the small BHs by $d$.

% The lengthscales characterizing the problem are illustrated in Fig. \ref{fig:physical_situation}.

%\begin{figure}
%\centre \includegraphics[width=.6\columnwidth]{physical_situation.pdf}
%\caption{Representation of the lengthscales of the problem considered. The distances have to satisfy eqs. \eqref{eq:condition_green} and \eqref{eq:condition_finite_size}}
%\label{fig:physical_situation}
%\end{figure}

We will focus on the inspiral of the two small BHs around each other, generating GWs potentially accessible in the LISA window \cite{Seoane:2013qna}. Such events were also considered as a probe of Kozai-Lidov oscillations generating eccentricity in the binary orbit \cite{Randall:2019sab,Randall:2017jop,Randall:2018nud}. In this subsection, we will show that there is a regime where one can perturbatively calculate the cubic Galileon corrections to the orbital parameters of the binary system, giving rise to a modification of the dynamics detectable in a GW signal. The rest of the article will be devoted to the derivation of this correction and its phenomenological implications.

 In this subsection we will concentrate on the scalar part of the action for illustrative purposes.
We take the matter action to be constituted of three sources with masses $m_0$, $m_1$ and $m_2$.
Furthermore, we will assume that the effective scalar couplings $\beta_\mathrm{eff}$ defined through Eqs.~\eqref{eq:def_vainshtein_radius}~-~\eqref{eq:matter_pp_action} are different for the small BHs and the supermassive one. Indeed, supermassive BHs (SMBH) are expected to be surrounded with accretion disks \cite{Abramowicz_2013, WAGONER2008828} with dynamical timescales $t_0$ of the order of the period of rotation of the disk. This gives rise to a dynamical evolution of the scalar $q_0=\mpl t_0^{-1}$. On the other hand, we assume that the binary system lives in a 'cleaner' environment and consequently the associated timescales of variation of the scalar $t_\mathrm{binary}$ are much longer, $t_\mathrm{binary} \gg t_0$ so that the scalar charge of the SMBH is much larger than the one of the binary system, $\beta_0 \gg \beta_\mathrm{binary}$.

 On flat spacetime and expanding to first order in $\pi / \mpl$, the scalar part of the matter action is
\begin{equation} \label{eq:matter_action_eff}
S_m = \int d^4 x \; \frac{\pi}{\mpl} T \; ,
\end{equation}
where the three bodies contribute as
\begin{equation}
T = - \beta_0 \; m_0 \delta^3(\vec{x}) - \beta_\mathrm{binary} \; m_1 \delta^3(\vec{x} - \vec{x}_1)- \beta_\mathrm{binary} \; m_2 \delta^3(\vec{x} - \vec{x}_2) \; ,
\end{equation}
up to relativistic corrections. The hierarchy $m_1, m_2 \ll m_0$ suggest the field decomposition $\pi = \pi_0 + \psi$, where $\pi_0$ is the spherically symmetric field generated by the central BH, eq. \eqref{eq:phi_SS} with $\beta_\mathrm{eff} = \beta_0$. $\psi$ represents a small perturbation of the scalar field generated by the two small BHs. In terms of this variable, the action reads
\begin{align} \label{eq:action_expanded}
\begin{split}
S[\pi] &= S[\pi_0] + \int d^4 x \left( -\frac{1}{2} (\partial \psi)^2 - \frac{1}{2\Lambda^3} \left[ (\partial\psi)^2 \Box \pi_0 + 2 \partial_\mu \psi \partial^\mu \pi_0 \Box \psi \right] \right. \\
 &- \left. \frac{1}{2\Lambda^3} (\partial\psi)^2 \Box \psi  +  \psi \frac{\tilde{T}}{\mpl}  \right) \; ,
\end{split}
\end{align}
where we have eliminated the linear terms with the equations of motion and introduced $\tilde T$ as
\begin{equation} \label{eq:tilde_T}
\tilde{T} = - \beta_\mathrm{binary} \; m_1 \delta^3(\vec{x} - \vec{x}_1)- \beta_\mathrm{binary} \; m_2 \delta^3(\vec{x} - \vec{x}_2) \; .
\end{equation}
By definition, the Vainshtein regime in the background of the central BH occurs  when the Galileon terms dominate the quadratic term $(\partial \psi)^2$ in the action. The behaviour of $\psi$ depends on which Galileonic term dominates in the action. Let us place ourselves close to one of the small bodies, say $m_1$. In this case, one can assume that the term cubic in $\psi$ dominates over the quadratic one and that we just recover the original cubic Galileon action for $\psi$. This means that $\psi$ is of the form
\begin{equation}
\psi = \left( \frac{\beta_\mathrm{binary} m_1 \Lambda^3 d }{2 \pi \mpl} \right)^{1/2} + C
\end{equation}
where $C$ is an irrelevant constant of integration, and $d$ is the distance to the source $1$. Indeed, taking the ratio of the term quadratic in $\psi$ to the cubic term yields \footnote{We have chosen the first quadratic term to illustrate our scaling. One obtains the same result with the second quadratic term $\partial_\mu \psi \partial^\mu \pi_0 \Box \psi$ if one first integrates it by parts in order to eliminate the higher derivatives on $\psi$.}
\begin{equation}
\frac{(\partial \psi)^2 \Box \pi_0}{(\partial \psi)^2 \Box \psi} \sim \frac{\Box \pi_0}{\Box \psi} \sim \left( \frac{d^3 \beta_0 m_0}{r^3 \beta_\mathrm{binary} m_1} \right)^{1/2}
\end{equation}
where we have used the scaling $\Box \pi_0 \sim \pi_0 / r^2$, $\Box \psi \sim \psi / d^2$. This means that sufficiently close to $m_1$, we can ignore the quadratic term and treat the central body only as a background field. In \cite{Kuntz:2019plo} it was noticed that this approximation is valid for the Sun-Earth-Moon system in a $P(X)$-type of theories and that it implied a violation of the Weak Equivalence Principle.  The maximal distance up to which one can neglect the quadratic terms in the action satisfies
\begin{equation} \label{eq:condition_finite_size}
\frac{d}{r_{V,1}} \ll \frac{r}{r_{V,0}}
\end{equation}
where $r_{V,0} \gg r_{V,1}$ are the Vainshtein radii (defined in equation \eqref{eq:def_vainshtein_radius}, replacing the masses and the couplings with the appropriate values) associated to the central and the first object in the absence of any other object, respectively. %As long as the size of the light bodies (their Scwharzchild radius $r_s$) respects condition \eqref{eq:condition_finite_size}, we can effectively treat them as point-masses.

On the other hand, if the distance between the light bodies is large enough, i.e
\begin{equation} \label{eq:condition_green}
\frac{d}{r_{V,1}} \sim \frac{d}{r_{V,2}}  \gtrsim \frac{r}{r_{V,0}}
\end{equation}
then the dominant term for $\psi$ in the action is the quadratic one. We will place ourselves in this physical situation. This can also be interpreted as the two BHs being outside their respective 'redressed' Vainshtein radius defined by
\begin{equation}
\tilde r_{V,\alpha}^3= \frac{\tilde \beta m_\alpha}{8 \pi \mpl \tilde\Lambda^3}, \ \alpha=1,2  \; .
\label{redr}
\end{equation}
In this equation, the redressed coupling and energy scale arise from canonically normalising the fluctuation $\psi$ and are of the form $\tilde \beta = \beta_\mathrm{binary} /\sqrt Z$, $\tilde \Lambda= \sqrt Z \Lambda $ where $Z = \Box \phi_0 /\Lambda^3 \sim (r_{V,0}/r)^3 $ is the factor in front of the kinetic term for $\psi$.
The condition $d \gtrsim  \tilde r_{V,1}$ expressing the fact the two BHs should be outside their redressed Vainshtein radius is perfectly equivalent to the condition \eqref{eq:condition_green}.

There is another physical process which we must take into account, namely tidal disruption of the binary system. When the tidal forces due to the large central BH are large enough, the binary constituents can no longer be held by their mutual attraction. This requires that
\begin{equation}
d \lesssim \left(\frac{m_1}{m_0} \right)^{1/3} r \; ,
\end{equation}
where we have used again $m_1 \sim m_2$. The domain of validity of our analysis is then
\begin{equation} \label{eq:condition_d}
\left(\frac{\beta_\mathrm{binary} m_1}{ \beta_0 m_0} \right)^{1/3} r \lesssim d \lesssim \left(\frac{m_1}{m_0} \right)^{1/3} r  \; .
\end{equation}
This condition enables us to calculate the cubic Galileon corrections to the orbital parameters  perturbatively.

\subsection{Three-body BHs: parametrisation of the problem} \label{subsec:param}

\begin{figure}[h!]
\center \includegraphics[width=.6\columnwidth]{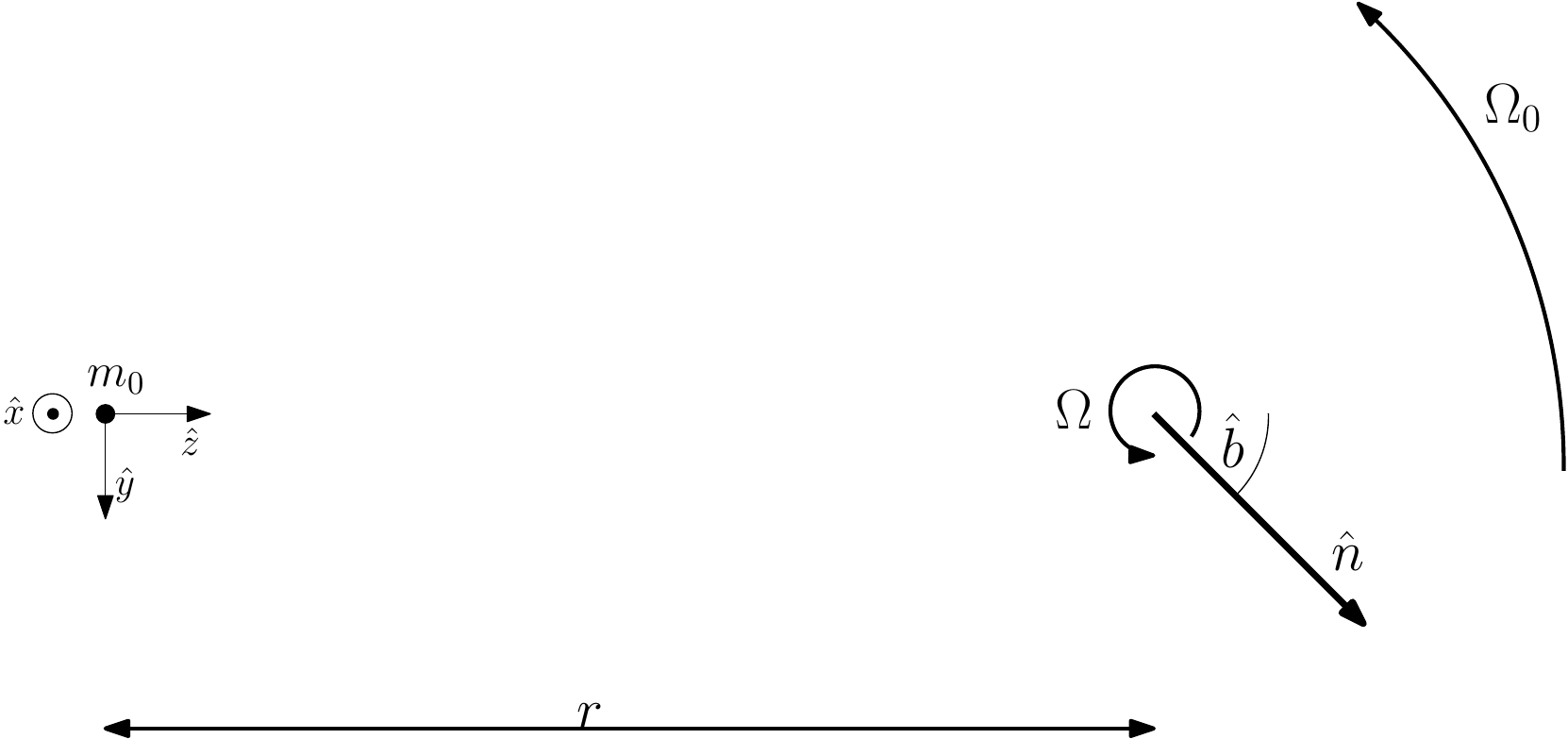}
\caption{Schematic representation of the length-scales of the problem at stake. The distances have to satisfy eqs. \eqref{eq:condition_green}, and $\hat n$ is the normal to the plane of rotation of the binary system formed by the two small BHs}
\label{fig:physical_situation}
\end{figure}

We are interested in the system of three BHs as described in section \ref{sec:three_body_problem} and will assume that the  BHs are described by a point-particle action.
For simplicity, we will assume all the motions to be circular, i.e the two small BHs are in a circular orbit of frequency $\Omega$ around their common barycentre. We call this motion the 'inner orbit', which is itself in circular orbit of frequency $\Omega_0$ around the massive BH. We call its motion the 'outer orbit'. This has been schematically illustrated in figure \ref{fig:physical_situation}. A more realistic treatment would necessitate to take  into account the eccentricity of the orbits which can grow to significant values in this kind of configurations \cite{Randall:2019sab,Randall:2017jop,Randall:2018nud}.

We perform the analysis in the frame co-rotating with the centre-of-mass of the binary. Denoting by $\vec d = \vec x_1 - \vec x_2$ and $\vec r$ the position of the centre-of-mass of the binary, we get
\begin{equation}
\vec x_1 \equiv \vec r + \vec d_1 = \vec r + (1-X) \vec d, \quad \vec x_2 \equiv \vec r + \vec d_2 = \vec r - X \vec d
\end{equation}
where $X = m_1/(m_1+m_2)$. The $z$ axis of the coordinate system is chosen to point from the distant massive object of mass $m_0$ at the origin to the centre-of-mass $\vec r$.

The orientation of the axis of rotation of the system is arbitrary and we choose to parametrise it with Euler angles $\hat a$, $\hat b$ and $\hat c$. This means that $\vec d$ is written as
\begin{equation}
\vec d = R_z(\hat c) R_x(\hat b) R_z(\hat a) (d \cos(\Omega t), d \sin(\Omega t), 0)^T \; ,
\end{equation}
where the $R_i$'s are rotation matrices around the $i$ axis. However, since the orientation of the $x$ and $y$ axes is defined up to an arbitrary rotation, we can fix $\hat c = 0$. In explicit coordinates $\vec d$ reads
\begin{equation}
\vec d = d
\begin{cases} \cos(\Omega t + \hat a) \\
  \cos(\hat b) \sin(\Omega t + \hat a) \\
  \sin(\hat b) \sin(\Omega t + \hat a)

\end{cases} \; ,
\end{equation}
which shows that $\hat a$ can be absorbed into a redefinition of the origin of time, so we set $\hat a = 0$ from now on.
Then the distance of object $1$ to the massive object $0$ is
\begin{equation}
\vert \vec x_1 \vert^2 = r^2 + 2 r d_1 \sin(\hat b) \sin(\Omega t) + d_1^2 \; ,
\end{equation}
where we recall that $r$ is the distance of the centre-of-mass to the massive object. The equations are the same for the object $2$ by just replacing $d_1$ by $d_2$.
 Before moving on, it is worth remarking that the condition \eqref{eq:condition_d} on the distances can be reformulated as a condition on the frequencies of the system using Kepler's third law. This reads
\begin{equation} \label{eq:bound_omega}
\Omega_0 \lesssim \Omega \lesssim \left( \frac{\beta_0}{\beta_\mathrm{binary}} \right)^{1/2} \Omega_0 \; .
\end{equation}
Hence the rotation rate of the binary system about itself is bounded from below, where tidal disruption takes place, and from above, where the two small BHs enter their respective Vainshtein radii. This has important consequences for the dynamics of the system as we shall see in section \ref{sec:inspiral}.

\section{Galileon propagation: the Green's function} \label{sec:Green}
In this section we  compute the propagation of the Galileon in the geometrical  background configuration. For this purpose we essentially need to compute the Green's function of the second order action in perturbations. This will enable us to compute the two-body energy and the power dissipated in scalar radiation.

Let us start from our base action \eqref{eq:base_action} and split the fields according to $g_{\mu \nu} = \eta_{\mu \nu} + h_{\mu \nu}$, $\pi = \pi_0 + \psi$ where $\pi_0$ is given by eq. \eqref{eq:phi_SS}. We do not consider the interactions between the scalar and the graviton since they will give subleading corrections to our results as we will argue below. This means that the action \eqref{eq:base_action} splits into the usual Einstein-Hilbert action and a cubic Galileon in flat spacetime coupled to matter through eq. \eqref{eq:matter_action_eff}. We now concentrate on the scalar part of the action, and as observed in Section \ref{sec:three_body_problem} we keep only terms quadratic in $\psi$ to obtain
\begin{equation}
S[\pi] = S[\pi_0] + \int d^4 x \left(  - \frac{1}{2\Lambda^3} \left[ (\partial\psi)^2 \Box \pi_0 + 2 \partial_\mu \psi \partial^\mu \pi_0 \Box \psi \right]   + \psi \frac{\tilde T}{\mpl} \right) \; ,
\end{equation}
where ${T}$ is given in eq. \eqref{eq:tilde_T}.

We rewrite the quadratic action for $\psi$ in spherical coordinates and integrate by parts in order to eliminate the higher derivatives on $\psi$. We find
\begin{equation}
S = S[\pi_0] + \int d^4x \frac{1}{2} \left[ K_{t} (\partial_t \psi)^2 - K_{r} (\partial_r \psi)^2 - K_{\Omega} (\partial_\Omega \psi)^2 \right] + \psi \frac{\tilde T}{\mpl}
\end{equation}
where the kinetic, angular and radial factors read
\begin{align}
K_t = 3 \left( \frac{r_*}{r} \right)^{3/2}, \qquad
K_r = 4 \left( \frac{r_*}{r} \right)^{3/2} \qquad \text{and} \qquad
K_\Omega = \left( \frac{r_*}{r} \right)^{3/2}
\end{align}
and the radius $r_*$
\begin{equation}
r_* = r_{V,0} = \left(\frac{\beta_0 m_0}{8 \pi \mpl \Lambda^3} \right)^{1/3}
\end{equation}
denotes the Vainshtein radius associated to the central mass.

In order to find the two-body energy and the power dissipated in scalar radiation, we have to find the Green's function associated to this quadratic operator. It is defined by
\begin{equation}
\left[- K_t(r) \partial_t^2 + \frac{1}{r^2} \partial_r \left(r^2 K_r \partial_r \right) + \frac{K_\Omega}{r^2} \nabla_\Omega^2 \right] G(x, x') = \delta^4(x - x')
\end{equation}
where $\nabla_\Omega^2 = \partial_\theta^2 + 1/\sin^2 \theta \partial_\phi^2$. It is worth mentioning  the following special boundary condition. We will take the field to vanish at the origin of coordinates where the massive object lies. Since the field goes as $r^{1/2}$ at the origin, this is consistent, contrary to the Newtonian problem where the field goes as $1/r$ but vanishes at infinity. We can obtain the field as
\begin{equation} \label{eq:field_green_fct}
\psi(x) = \int d^4x' G(x, x') \left( - \frac{ \tilde T(x')}{\mpl} \right)
\end{equation}

In the static case, by substituting  the field in the quadratic action, we can obtain the two-body energy, defined as $\int dt E = - \Re(S_\mathrm{cl})$
\begin{equation} \label{eq:energy_green_fct}
- \Re(S_\mathrm{cl})=\int dt E = \frac{1}{2} \int d^4x d^4x' G(x, x') \frac{\tilde T(x)}{\mpl} \frac{\tilde T(x')}{\mpl} \; .
\end{equation}
We will follow  Refs. \cite{Andrews_2013,deRham:2012fg,Chu:2012kz} in order to calculate the Green's function.
For static configurations, we will subtract an infinite term from the Green's function which is space-independent and does not contribute to the energy as the supports of the
matter distributions do not intersect.

 By introducing the rescaled variable $\vec u = \frac{\vec r}{r_*}$, we rewrite the previous equation as
\begin{equation}
\left( -3 r_*^2 \partial_t^2 + 4 \partial_u^2 + \frac{2}{u} \partial_u + \frac{\nabla_\Omega^2}{u^2} \right) G(\vec u, t ; \vec u', t') = \frac{\delta^3(\vec u - \vec u') \delta(t-t')}{r_*}\;.
\end{equation}
We further decompose the Green function in a Fourier and spherical harmonics basis as
\begin{equation} \label{eq:green_fourier}
G(\vec u, t ; \vec u', t') = \int \frac{d\omega}{2\pi} e^{-i \omega (t-t')} \sum_{l=0}^{\infty} R_{lm \omega}(u, u') \sum_{m=-l}^l Y_l^m[\theta, \phi] \bar Y_l^m[\theta', \phi']\;.
\end{equation}
Using the resolution of the identity
\begin{equation}
\sum_{l,m} Y_l^m[\theta, \phi] \bar Y_l^m[\theta', \phi'] = \delta( \cos(\theta) - \cos(\theta') ) \delta(\phi - \phi')
\end{equation}
one easily finds the equation for the mode function $R_{lm}$
\begin{equation} \label{eq:diffeq_R}
\left( \partial_u^2 + \frac{1}{2u} \partial_u + \frac{3}{4} \tilde \omega^2 - \frac{l(l+1)}{4 u^2} \right) R_{lm \omega}(u,u') = \frac{\delta(u - u')}{4 r_* \sqrt{u}}
\end{equation}
where $\tilde \omega = r_* \omega$.

The general continuous solution of eq. \eqref{eq:diffeq_R} is
\begin{equation}
R(u,u') = A R^1(u) R^1(u') + B R^2(u) R^2(u') + C R^1(u_<) R^2(u_>) + D R^1(u_>) R^2(u_<)
\end{equation}
where we have omitted the index $l,m, \omega$ for clarity and the constants $A,B,C,D$ are to be fixed by the normalization of the mode functions and the boundary conditions. Here we have introduced the notation $u_> = \mathrm{max}(u, u')$ and $u_< = \mathrm{min}(u, u')$, and the homogeneous solutions are given by the Bessel functions
\begin{align}
\begin{split}
R^1(u) &= \mathcal{N} u^{1/4} J_\nu \left(\sqrt{\frac{3}{4}} u \tilde \omega \right) \\
R^2(u) &= \mathcal{N} u^{1/4} J_{-\nu}\left(\sqrt{\frac{3}{4}} u \tilde \omega \right)\;,
\end{split}
\end{align}
where $\mathcal{N}$ is a normalization constant and $\nu = (2l+1)/4$. Integrating eq. \eqref{eq:diffeq_R} for $u$ close to $u'$, we get
\begin{equation}
(D-C)W = \frac{1}{4r_* \sqrt{u}}
\end{equation}
where
\begin{equation}
W = R^{1'} R^2 - R^1 R^{2'} = \frac{2 \mathcal{N}^2 \sin(\nu \pi)}{\pi \sqrt{u}}
\end{equation}
is the Wronskian of the two homogeneous solutions. We choose
\begin{equation}
\mathcal{N} = \sqrt{\frac{\pi}{8r_* \sin(\nu \pi)}}
\end{equation}
such that $D-C = 1$.

We next determine the constants from the boundary conditions. We require the flux to be purely outgoing at infinity, which corresponds to taking the retarded Green's function. On the other hand, the boundary condition at the origin can be fixed by the following observation. Consider the field produced by a variation $m_0 \rightarrow m_0 + \delta m_0$ of the central mass. From eq. \eqref{eq:phi_SS} (with a field vanishing at the origin), it is
\begin{equation}
\delta \pi_0(r) = \frac{\beta_0 \delta m_0}{8\pi \mpl r_*} \left( \frac{r}{r_*} \right)^{1/2}
\end{equation}
On the other hand, from eq. \eqref{eq:field_green_fct} with a source replaced by $\delta T = - \beta_0 \delta m_0 \delta^3(\vec x)$, we have
\begin{align}
\begin{split}
\delta \pi_0(t, \vec x) &= - \frac{1}{\mpl} \int d^4x' G( x, x') \delta T( x') \\
&= \frac{\beta_0 \delta m_0}{\mpl} \int dt' G(\vec r, t ; \vec 0, t')\;.
\end{split}
\end{align}
By equating these two equations, we find the boundary condition at the origin
\begin{equation}\label{eq:BC_origin}
\mathrm{lim}_{\omega \rightarrow 0} \sum_l \frac{2l+1}{4\pi} P_l(\cos(\theta)) R_{lm\omega}(u,0) = \frac{\sqrt{u}}{8\pi r_*}\;,
\end{equation}
where $P_l$ represent the Legendre polynomials, and we have used the following identities
\begin{equation}
Y_l^m(0,\phi) = \sqrt{\frac{2l+1}{4\pi}} \delta_{m0} \; , \quad Y_l^0(\theta, \phi) = \sqrt{\frac{2l+1}{4\pi}} P_l(\cos(\theta))\;.
\end{equation}

Let us examine the asymptotic behaviour for $l>0$ first. From the behaviour of the Bessel's functions at the origin
\begin{equation}
J_\nu(z) \sim \frac{1}{\Gamma(\nu+1)} \left( \frac{z}{2} \right)^\nu
\end{equation}
we immediately deduce that $B=D=0$ for $l \geq 1$ in order for the Green's function to be continuous at the origin. The solution is now
\begin{equation}
R(u,u') = A R^1(u) R^1(u') - R^1(u_<) R^2(u_>)\;.
\end{equation}
The constant $A$ is fixed by requiring the flux to be outgoing at infinity. Indeed, by rewriting the Bessel functions in terms of the two Hankel functions
\begin{align}
\begin{split}
J_\nu &= \frac{H_\nu^{(1)} +H_\nu^{(2)} }{2} \\
J_{-\nu} &= \frac{1}{2} \left( H_\nu^{(1)}(1-i \tan(\nu \pi)) - H_\nu^{(2)}(1+i \tan(\nu \pi)) \right)
\end{split}
\end{align}
and using the asymptotic behaviour at infinity
\begin{align} \label{eq:asymptotic_bessel}
\begin{split}
H_\nu^{(1)}(z) &\sim \sqrt{\frac{2}{\pi z}} e^{i\left(z-\nu \frac{\pi}{2} - \frac{\pi}{4} \right)} \\
H_\nu^{(2)}(z) &\sim \sqrt{\frac{2}{\pi z}} e^{-i\left(z-\nu \frac{\pi}{2} - \frac{\pi}{4} \right)}
\end{split}
\end{align}
the condition that the flux is purely outgoing imposes
\begin{equation}
A = - (1+i \tan(\nu \pi))
\end{equation}

Let us now examine the $l=0$ case. In this case, $R^2$ is not divergent any more at the origin but takes a finite value,
\begin{equation}
R^2(0) = \frac{1}{\Gamma(3/4)} \left( \frac{1}{8 \sqrt{3} \tilde \omega} \right)^{1/4} \sqrt{\frac{\pi}{r_*}}\;.
\end{equation}
The solution when one of the points is taken to be the origin is
\begin{equation}
R(u,0) = R^2(0) \left( B R^2(u) + D R^1(u) \right)\;.
\end{equation}
We can now use eq. \eqref{eq:BC_origin} by noticing that $R_{lm \omega}(u,0) = 0$ for $l > 0$, which gives
\begin{equation}
\mathrm{lim}_{\omega \rightarrow 0} R(u,0) = \frac{\sqrt{u}}{2 r_*}\;.
\end{equation}
This imposes $D = 1$ (so $C=0$). The $B$ coefficient multiplies a power-law divergent term which we simply subtract as it does not depend on the variables $u, u'$.  $B$ is  left undetermined here.
To find $A$ and $B$, let us rewrite the solution when one of the endpoints is taken to infinity, say $u'$,
\begin{equation}
R(u,u') = A R^1(u) R^1(u') + R^2(u)(B R^2(u') + R^1(u')) \; ,
\end{equation}
so that in order to have a purely outgoing flux, one should impose
\begin{align}
\begin{split}
A &= 0 \\
B &= \frac{1}{1+i}\;.
\end{split}
\end{align}
In conclusion, the mode functions are
\begin{equation} \label{eq:mode_functions}
\begin{array}{rclc}
R(u,u') &=& R^1(u_>) R^2(u_<) + \frac{1}{1+i} R^2(u) R^2(u') & \text{for }l=0 \\
& & & \\
R(u,u') &=& - R^1(u_<) R^2(u_>) - (1+i \tan(\nu \pi)) R^1(u) R^1(u') & \text{for }l>0  \;.
\end{array}
\end{equation}
In this way the Green's function \eqref{eq:green_fourier} is completely characterised.

\section{Static limit: $\omega \rightarrow 0$} \label{sec:static}

\subsection{The two-body energy}
Next, we shall discuss the static limit (i.e, the limit $\omega \rightarrow 0$) of the Green's function which will give rise to the two-body scalar energy. Let us focus on $l>0$ first. Then it is easy to see that the only nonzero part of $R$ is
\begin{align}
\begin{split}
R(u,u') &= - R^1(u_<) R^2(u_>) \\
&= - \frac{1}{2 (2l+1) r_*} u_<^{(l+1)/2} u_>^{-l/2}\;,
\end{split}
\end{align}
which gives upon using eq. \eqref{eq:green_fourier} and a choice of axis for $ \vec u '$ such that $\theta ' = 0$
\begin{equation}
\int dt' G(\vec u, t ; \vec u', t') \supset \frac{1}{8 \pi r_*} \left( u_<^{1/2} - \frac{u^{1/2} u'^{1/2}}{\vert \sqrt{u} \hat u - \sqrt{u'} \hat u' \vert} \right) \; ,
\end{equation}
where $\hat u = \vec{u} / u$, and we have used the Legendre polynomial identity
\begin{equation}
\frac{1}{\vert \vec x - \vec x' \vert} = \sum_{l=0}^{\infty} \frac{x'^l}{x^{l+1}} P_l(\cos\theta) \; ,
\end{equation}
where $x = \vert \vec x \vert$, $x' = \vert \vec x' \vert$ and $\theta$ is the angle between $\vec x$ and $\vec x'$.
Ignoring the divergent $R^2 R^2$ contribution as before, the $l=0$ term is similarly obtained and one finally gets for the static Green's function
\begin{equation}
\int dt' G(\vec u, t ; \vec u', t') = \frac{1}{8 \pi r_*} \left( - \frac{\rho \rho'}{\vert \vec{\rho} - \vec{\rho}' \vert}  + \rho + \rho' \right)
\end{equation}
where $\vec \rho = \sqrt{u} \hat u$.

We shall use this Green's function to find the energy. There are two types of terms in eq. \eqref{eq:energy_green_fct}: the self-energy terms and the $1-2$ interaction term. We will ignore the self-energies since these terms contribute only when finite-size effects are taken into account. In any case the evaluation of the Green's function at coincident points is not well defined because eq. \eqref{eq:condition_green} is not satisfied.
The interaction term gives the energy
\begin{equation} \label{eq:energy}
E = \frac{\beta_\mathrm{binary}^2 m_1 m_2}{2 \pi \mpl^2 r_*} \left( \rho_1 + \rho_2 - \frac{\rho_1 \rho_2}{\vert \vec{\rho}_1 - \vec{\rho}_2 \vert} \right)
\end{equation}

Let us now come back to Figure \ref{fig:physical_situation} and give some order-of-magnitude estimates. If $d \lesssim  r$, the last term in the energy dominates
\begin{equation}
\frac{\rho_1 \rho_2}{\vert \vec{\rho}_1 - \vec{\rho}_2 \vert} \sim \sqrt{\frac{r}{r_*}} \frac{r}{d}\;.
\end{equation}
Then the ratio of this energy to the Newtonian energy $E_N \sim m_1 m_2/( \mpl^2 d)$ is
\begin{equation}
\frac{E}{E_N} \sim \beta_\mathrm{binary}^2 \left( \frac{r}{r_*} \right)^{3/2}\;.
\end{equation}
We thus see that the energy is screened with respect to the distance to the massive object. There is no factor involving the distance between the two bodies $d$. We will now evaluate precisely the scalar energy in the case of a circular orbit.

\subsection{The case of circular orbits}
In this subsection we will evaluate the two-body scalar energy for circular orbits.
Using the notations of Section \ref{subsec:param}, we find, when $\vert \vec d \vert \lesssim \vert \vec r \vert$,
\be
\vec \rho_{1,2}\simeq \frac{1}{r^{1/2}r_* ^{1/2}} \vec r + \frac{1}{r^{1/2}r_* ^{1/2}} (\vec r_{1,2}-\vec r) - \frac{1}{2} ((\vec r_{1,2}-\vec r).\vec r) \frac{\vec r}{r_*^{1/2} r^{5/2}}\;,
\ee
which imply that to leading order
\begin{align}
\rho_{1,2} &\simeq \frac{r^{1/2}}{r_*^{1/2}} \\
\vert \vec \rho_1 -\vec \rho_2\vert &\simeq \frac{d}{r_*^{1/2}r^{1/2}} \sqrt {1 - \frac{3}{4} \sin^2 \hat b \sin^2 \Omega t} \; .
\end{align}
As stated just above, the leading term in the two-body energy \eqref{eq:energy} is the last one,
\begin{equation}
E \simeq - 4 G \beta_\mathrm{binary}^2 \frac{m_1 m_2}{d \sqrt{1 - \frac{3}{4} \sin^2 \hat b \sin^2 \Omega t }} \left( \frac{r}{r_*} \right)^{3/2} \; .
\end{equation}

By averaging over time, we see that the effect of the scalar energy, when added to the gravitational energy $E_\mathrm{gr} = - G m_1 m_2/d$, is just a renormalization of Newton's constant $G \rightarrow G_\mathrm{eff}$, where the effective gravitational coupling constant takes the form
\begin{equation}
G_\mathrm{eff} = G \left(1 + \alpha \beta_\mathrm{binary}^2 \left( \frac{r}{r_*} \right)^{3/2}  \right) \; ,
\end{equation}
with $\alpha$ denoting a numerical quantity given by
\begin{equation}
\alpha = \frac{4}{\pi} \left[ K \left( \frac{3 \sin^2 \hat b}{4} \right) + \frac{2}{4 - 3 \sin^2 \hat b} K \left( \frac{3 \sin^2 \hat b}{3 \sin^2 \hat b - 4} \right) \right] \; ,
\end{equation}
and $K$ is the complete elliptic integral of the first kind,
\begin{equation}
K(m) = \int_0^\frac{\pi}{2} \frac{d u}{\sqrt{1 - m \sin^2 u}} \; .
\end{equation}
This result corroborates  the expectation that the two-body scalar energy is a $\beta^2$ correction screened by the Vainshtein factor $(r/r_*)^{3/2}$.

\section{Dissipated power} \label{sec:diss_power}
In this section we  compute the dissipated power in scalar radiation due to the presence of the cubic Galileon. The power emitted in the tensor sector will follow the usual quadrupole formula and hence we will focus on the scalar sector.

\subsection{Energy-momentum tensor}
The total energy-momentum tensor splits into gravitational and scalar contributions,
\begin{equation}
T_{\mu \nu} = T^\pi_{\mu \nu} + T^g_{\mu \nu} \; ,
\end{equation}
where $T^g_{\mu \nu}$ is the usual Landau-Lifschitz pseudo-tensor, and $T^\pi_{\mu \nu} = - 2 / \sqrt{-g} \delta S_\pi/\delta g^{\mu \nu}$, where $S_\pi$ is the scalar part of the action. Far from matter sources the total energy-momentum tensor is conserved, which allows to find the power lost into radiation by integrating it over a distant sphere of radius $\mathcal{R}$ centred on the system,
\begin{equation}
P = \int d^2 S T_{0i}n^i = \mathcal{R}^2 \int d^2 \Omega T_{0r} \; ,
\end{equation}
where $n^i$ is the outward pointing vector of the sphere. The Landau-Lifschitz pseudo-tensor will give rise to the usual quadrupole formula at lowest order in the post-Newtonian expansion, so there remains only to find the scalar dissipated power. The scalar energy-momentum tensor calculated from the action \eqref{eq:base_action} reads
\begin{align}
\begin{split} \label{eq:EMT}
T^\pi_{\mu \nu} &= \partial_\mu \pi \partial_\nu \pi - \frac{1}{2} g_{\mu \nu} (\partial \phi)^2 + \frac{1}{\Lambda^3} \left( \partial_\mu \pi \partial_\nu \pi \square \pi + \frac{1}{2} g_{\mu \nu} \partial_\alpha \big( (\partial \pi)^2 \big ) \partial^\alpha \pi \right. \\
&- \left. \frac{1}{2} \big [ \partial_\mu \big( (\partial \pi)^2 \big) \partial_\nu \pi + \mathrm{sym} \big ]  \right) \; .
\end{split}
\end{align}
Splitting the field $\pi = \pi_0 + \psi$ as in Eq. \eqref{eq:action_expanded}, one can collect the terms quadratic in $\psi$ in the energy-momentum tensor. The linear terms average to zero in time in the dissipated power. As emphasised in Section \ref{sec:setup}, the dominant terms will be the quadratic ones coming from the Galileon term. By neglecting angular and time total derivatives, which, once again, will average to zero in the dissipated power, one finds the $0r$ part of the scalar energy-momentum tensor
\begin{equation}
T^\pi_{0r} = \frac{4  \pi_0'}{\Lambda^3 \mathcal{R}} \partial_t \psi \partial_r \psi.
\end{equation}
For a wave travelling far from the massive objects, one has $\partial_r \psi = - \partial_t \psi / c_r$ where $c_r = \sqrt{3}/2$ is the radial propagation speed. Using Eq. \eqref{eq:phi_SS}, this gives
\begin{equation} \label{eq:power}
P = \frac{8}{\sqrt{3}} \mathcal{R}^{1/2} r_*^{3/2} \int d^2 \Omega (\partial_t \psi)^2\;.
\end{equation}

\subsection{Dissipated power}

%Let us now find the field radiated at infinity. We will assume a slow motion of the small bodies $v \ll 1$ and expand all quantities in powers of the velocity. As we deal with scalar fields, there will generally be monopole and dipole radiation on top of the quadrupole one. Generically, each $n$-pole is suppressed by $v^{2n}$ in the velocity expansion with respect to the monopole.
%
%Before dealing with the calculations, let us review briefly the generic results obtained in the simpler setup of scalar-tensor theories {\bf CITATION}. In these theories, the lowest-order monopole is simply proportional to the total mass of the system and so does not radiate since it is conserved. On the other hand, if the scalar charges of the two objects are different, then there will be a dipole radiation enhanced by $1/v^2$ with respect to the quadrupole, and suppressed by the difference of scalar charges of the two objects. If the two objects share the same scalar charge, then by conservation of momentum there is no dipole radiation at lowest order.
%
%Let us now return to the Galileon case and assume that the two objects share different scalar charges, $\beta_1 \neq \beta_2$. The massive central object is assumed to have another scalar charge $\beta_0$. We will see that the lowest order contribution to the monopole is proportional to the difference of scalar charges of the two objects, $\beta_1 - \beta_2$.
%
%{\bf TODO : I IMPLICITELY ASSUME HERE THAT THE MOTION OF THE BINARY SYSTEM AROUND THE CENTRAL BLACK HOLE IS SLOW...}

Let us now find the power radiated at infinity, which reduces to finding $\psi(\vec x,t)$ at large distance from the source. In order to find the lowest-order power emitted we will assume that the two objects follow non-relativistic trajectories. The trace of their energy-momentum tensor is given by
\begin{equation}
T = - \beta_\mathrm{binary} m_1 \delta^3(\vec{x} - \vec{x}_1)- \beta_\mathrm{binary} m_2 \delta^3(\vec{x} - \vec{x}_2)\;.
\end{equation}
If one wanted to find higher-order corrections to the dissipated power, one would need to add relativistic corrections to this equation, but this is unnecessary for the lowest-order power as we argue below. By using eqs. \eqref{eq:field_green_fct} and \eqref{eq:green_fourier}, one finds
\begin{align}
\begin{split} \label{eq:field}
\psi(\vec x, t) &= \frac{\beta_\mathrm{binary}}{\mpl} \int dt' \frac{d\omega}{2\pi} e^{-i \omega(t-t')} \sum_{l,m} Y_l^m(\theta, \phi) \\
& \times \left[ m_1 R_{l\omega}(\mathcal{R}, \vert \vec r + \vec d_1(t') \vert) \bar Y_l^m (\theta_1(t'), \phi_1(t')) + (1 \leftrightarrow 2) \right] \; ,
\end{split}
\end{align}
where $\mathcal{R}, \theta, \phi$ are the coordinates of the distant sphere of integration (at large distance, we can set the centre-of-mass of the binary system and the position of the large mass $m_0$ to the same point), and $\theta_1(t), \phi_1(t)$ are the polar coordinates of the object 1 when taking the central mass to be the origin of the coordinates.

Equation \eqref{eq:field} represents the field in the frame corotating with the system. It is easy to find the field in the static frame attached to the central BH: one just has to replace $\phi_i(t') \rightarrow \phi_i(t') + \Omega_0 t'$ (where $\Omega_0$ is the angular speed of rotation of the barycentre of the system around the massive black hole)  in the spherical harmonics, which adds a time dependent factor $e^{-i m \Omega_0 t'}$ to the total expression.

We can then use that, for a $T$-periodic function $f$, the Fourier series becomes
\begin{equation}
\int dt' e^{i \omega t'} f(t') = 2 \pi \sum_{n \in \mathbb{Z}} c_n(f) \delta(\omega - n \Omega), \quad c_n(f) = \frac{1}{T} \int_0^T dt' f(t') e^{i n \Omega t'} \; .
\end{equation}
We find that
\begin{align}
\begin{split}
\psi(\vec x, t) &= \beta_\mathrm{binary} \frac{ (m_1+m_2)}{\mpl} \sum_{n \in \mathbb{Z}} e^{-i (n \Omega + m \Omega_0) t} \sum_{l,m} Y_l^m(\theta, \phi) (c_{n}(f_1^{lm}) + (1 \leftrightarrow 2)) \\
f_i^{lm}(t) &= X_i R_{l\omega}(\mathcal{R}, \vert \vec r + \vec d_i(t') \vert) \bar Y_l^m (\theta_i(t'), \phi_i(t')) \; ,
\end{split}
\end{align}
where $X_1 = X = m_1/(m_1+m_2)$, $X_2=1-X$ and $\omega = n \Omega + m \Omega_0$.
Finally, using eq. \eqref{eq:power} and integrating over a sphere, the total power is
\begin{equation} \label{eq:power_cn}
P = \frac{8}{\sqrt{3}} \beta_\mathrm{binary}^2 \mathcal{R}^{1/2} r_*^{3/2} \frac{(m_1+m_2)^2}{\mpl^2} \sum_{n \in \mathbb{Z}} \sum_{l,m} (m \Omega_0 + n \Omega)^2  \vert c_n(f_1^{lm}) + c_n(f_2^{lm}) \vert^2\;.
\end{equation}
At this point we can check the convergence of the different sums. The expansion in $l$ involves the mode fonction $R_{l\omega}(\mathcal{R}, x_i)$ (denoting $ x_i = \vert \vec x_i \vert = \vert \vec r + \vec d_i \vert$). More precisely, it involves a Bessel function evaluated in $x_i \Omega$.  Using the bound on the speed of rotation of the binary system, Eq. \eqref{eq:bound_omega}, we get that:
\begin{equation} \label{eq:r_omega}
r \Omega \lesssim \left( \frac{\beta_0}{\beta_\mathrm{binary}} \right)^{1/2} (r \Omega_0) \; .
\end{equation}
On the one hand, $r \Omega_0 = v \ll 1$ is the speed of the binary system  with respect to the central BH. On the other hand, we assume $\beta_0 / \beta_\mathrm{binary} \gg 1$ for our perturbative computation to be valid. We will now further assume that $r \Omega \lesssim 1$ which means that the ratio of scalar charges is moderately large.
Consequently, we can evaluate the Bessel function for small arguments and each order in $l$ is  suppressed by $(r \Omega)^{(l+1)/2}$ (apart from the monopole whose scaling is different as we show below).
Terms in the $n$-expansion  arise by expanding the distance $x_i \simeq r + d_i \sin(\hat b) \sin(\Omega t) + \dots$ in powers of $d/r$ leading to  Fourier coefficients from the corresponding power of $\sin(\Omega t)$. Thus, the expansion in $n$ corresponds to an expansion in $d/r$.

\subsection{Monopole}

The monopole ($l=m=0$) term is simple as the spherical harmonics are given by $Y_0^0 = 1/\sqrt{4\pi}$. From eq. \eqref{eq:mode_functions}, the mode function $R_{0\omega}$ for large value of $\mathcal{R}$ is given by
\begin{align}
\begin{split}
R_{0\omega}(\mathcal{R}, x_i) &= R^2(x_i)\left( R^1(\mathcal{R}) + \frac{1}{1+i} R^2(\mathcal{R}) \right) \\
&= \frac{\sqrt{2 \pi}(1-i)}{8 \cdot 3^{1/4} r_*^{3/2} \omega^{1/2}} \left( \frac{ x_i}{\mathcal{R}} \right)^{1/4} e^{i \sqrt{3} \omega \mathcal{R}/2 - i 3\pi/8} J_{-1/4}\left( \frac{\sqrt{3}}{2} x_i \omega \right) \; ,
\end{split}
\end{align}
where we have used the behaviour \eqref{eq:asymptotic_bessel} at infinity.
One can then Taylor expand the Bessel function as $x_1 \simeq r + d_1 \sin(\hat b) \sin(\Omega t) + \dots$ and  $d \lesssim r$. As the scalar field is real  only the even powers of  $(d/r)^{2k}$ appear in the sum of the Fourier coefficients $c_n(f_1) + c_n(f_2)$. They arise as $d$ accompanies  $ \sin(\Omega t)$ in the expansion and  terms of order $d^{2k}$ will only give contributions to harmonics of order  $n \leq 2k$.

Since the constant $n=0$ term does not give any power in monopole (see eq. \eqref{eq:power_cn}), the lowest-order contribution is given by $n=2$ (so $\omega = 2 \Omega$) with at second order in $d/r$
\begin{align}
\begin{split} \label{eq:cn_monopole}
c_2(f_1^{00}) + c_2(f_2^{00}) &=  X(1-X) J_{7/4}\left( \sqrt{3} r \Omega \right) \sin^2(\hat b) \\
&\times (1-i)e^{ i \sqrt{3} \Omega \mathcal{R}/2 - i 3\pi/8} \frac{3^{3/4}}{2^7} \frac{ r^{1/4} (d \Omega)^2}{\mathcal{R}^{1/4} \Omega^{1/2} r_*^{3/2}} \; .
\end{split}
\end{align}
 An important observation is that this second-order expansion depends on $(d\Omega)^2$ and not on $(d/r)^2$. Since $(d\Omega)^2 / (d/r)^2 = (r \Omega)^2 \lesssim 1$ as we have assumed below Eq. \eqref{eq:r_omega}, this makes the monopole more suppressed than the other modes which depend on $d/r$ as we confirm below.
%{\bf An important observation is that this second-order expansion depends on $(d\Omega)^2$ and not on $(d/r)^2$. Since in our approximation scheme $d/r$ becomes of order $d/r \gtrsim (\beta/\beta_0)^{1/3} 1/50$, while $d \Omega \sim 10^{-3}$, this makes the monopole suppressed than the other modes as long as $\beta_0\lesssim 10^4 \beta$ . In the following, we shall consider cases where the ratio between the coupling constant is smaller than this and we shall neglect the monopole channel of dissipation.  }
Another important point to notice is that relativistic corrections to the energy-momentum tensor as well as couplings between gravity and the field $\psi$ would correct this expression with terms scaling precisely as $(d \Omega)^2$. Thus, since a precise evaluation of the monopole would require these terms as well, we shall neglect the monopole channel of dissipation and we move on to the calculation of the dipole.

\subsection{Dipole}

The dipole term is defined by $l=1$, $m =0, \pm 1$. The mode function $R_{1 \omega}$ is given by
\begin{align}
\begin{split} \label{eq:mode_fonction_dipole}
R_{1\omega}(\mathcal{R}, x_i) &= - R^1(x_i)\left( R^2(\mathcal{R}) + (1-i) R^1(\mathcal{R}) \right) \\
&= - \frac{\sqrt{2 \pi}}{4 \cdot 3^{1/4} r_*^{3/2} \omega^{1/2}} \left( \frac{ x_i}{\mathcal{R}} \right)^{1/4} e^{i \sqrt{3} \omega \mathcal{R}/2 - i 5\pi/8} J_{3/4}\left( \frac{\sqrt{3}}{2} x_i \omega \right) \; ,
\end{split}
\end{align}

 We split up the computation between $n=0$, which represents the dipolar emission of the centre-of-mass due to its motion around the central mass and $n=2$ which contains also dipolar emission from the motion of the small masses themselves around their centre-of-mass.

\subsubsection{$n=0$ term} \label{sec:dip_n0}

The $n=0$ term corresponds to replacing $x_i \rightarrow r$ at lowest order in the $d/r$ expansion. Only the $m = \pm 1$ modes radiate (see eq. \eqref{eq:power_cn}). The relevant spherical harmonics are given by
\begin{equation}
Y_1^1(\theta, \phi) = - \bar Y_1^{-1}(\theta, \phi) = - \sqrt{\frac{3}{8 \pi}} \sin \theta e^{i \phi} \; .
\end{equation}
Moreover, with the replacement $x_i \rightarrow r$, i.e positions are evaluated at the barycentre of the system, one has $\theta_i = \phi_i = 0$. Using the expression for the mode function \eqref{eq:mode_fonction_dipole} it is easy to find
\begin{equation} \label{eq:dipole_n0term}
P_{n=0} = \frac{\beta_\mathrm{binary}^2}{4} J_{3/4}^2\left( \frac{\sqrt{3}}{2} r \Omega_0 \right) \frac{(m_1+m_2)^2}{\mpl^2} \left( \frac{r}{r_*} \right)^{1/2} \frac{\Omega_0}{r_*}
\end{equation}

Let us evaluate the order-of-magnitude of this expression for $r \Omega_0 \lesssim 1$. We find
\begin{equation}
P_{n=0} \sim \frac{\beta_\mathrm{binary}^2}{G} \frac{1}{(\Omega_0 r_*)^{3/2}} \left( \frac{m_i}{m_0} \right)^2 v^8
\end{equation}
where we have used the variable $v$ defined as $v=(G m_0 \Omega_0)^{1/3}$, which scales as a velocity in the post-Newtonian expansion and is related to the radius using Kepler's third law
\begin{align}
r = \left( \frac{G m_0}{\Omega_0^2} \right)^{1/3}.
 \equiv G m_0 v^{-2}
\end{align}
Comparing to the GR quadrupole power loss of a point-particle in circular orbit, $P_\mathrm{GR} \sim (m_i/m_0)^2 v^{10} / G$, we see that the dipole emission is enhanced by two powers of the velocity (as expected), but is suppressed by the Vainshtein factor $1/(\Omega_0 r_*)^{3/2}$, in line with the findings of Refs. \cite{deRham:2012fg,Dar:2018dra,deRham:2012fw}. This factor is extremely small as long as the Vainshtein radius is much larger than the size of the system, more precisely the wavelength of the emitted radiation. This implies that the  Vainshtein screening is  quite efficient for the global motion of the barycentre around the central mass, even if the dipolar power is generically enhanced by two powers of the velocity.

\subsubsection{$n=2$ term}

As explained above, only the even $n$ can contribute to the power. Let us first concentrate on the $m=0$ term, which is the most interesting because it does not involve the rotation $\Omega_0$ around the massive black hole and so represents a power loss for the binary system itself. The expression for the $l=1,m=0$ spherical harmonic is
\begin{equation}
Y_1^0(\theta, \phi) = \sqrt{\frac{3}{4 \pi}} \cos(\theta) \; .
\end{equation}

Let us choose $\theta$ and $\phi$ such that the barycentre of the system lies in the $\theta=\pi/2$ plane and such that this plane is perpendicular to the $x$ axis (of unit vector $\hat x$). Then trigonometry gives
\begin{equation}
\cos \theta_i = \frac{\vec d_i \cdot \hat x}{\sqrt{r^2 + (d_i \cdot \hat x)^2}} \; ,
\end{equation}
where $\theta_i$ is the $\theta$ angle of the object $i=1,2$. With this in hand, one can compute the Fourier coefficient
\begin{align}
\begin{split}
c_2(f_1^{10}) + c_2(f_2^{10}) &= \left(  X(1-X) J_{3/4}\left( \sqrt{3} r \Omega \right) \sin(\hat b) + \mathcal{O}(r \Omega) \right) \\
&\times i e^{ i \sqrt{3} \Omega \mathcal{R} - i 5\pi/8} \frac{3^{1/4}}{32 r_*^{3/2} \Omega^{1/2}} \left( \frac{r}{\mathcal{R}} \right)^{1/4} \frac{ d^2}{r^2} \; ,
\end{split}
\end{align}
where we have kept the leading term in $r \Omega$, and the $X(1-X)$ prefactor comes from replacing $d_i \rightarrow X_i d$. We thus confirm our previous claim: comparing to eq. \eqref{eq:cn_monopole}, we see that this term involves the small parameter $d/r$ instead of the much smaller one $d \Omega$. Thus, the dipole will dominate the dissipated power.

Replacing in the expression of the Fourier coefficients for the power (by symmetry, for $m=0$ the positive and negative $n$ give  the same contribution to the power) we find
\begin{equation}
P_{n=2, m=0} = \frac{\beta_\mathrm{binary}^2}{16} \frac{(m_1+m_2)^2}{\mpl^2} \frac{\Omega d^4}{r^{7/2} r_*^{3/2}} X^2(1-X)^2  J_{3/4}^2\left( \sqrt{3} r \Omega \right) \sin^2(\hat b)
\end{equation}

Introducing as for the $n=0$ term the variable $u=(G (m_1+m_2) \Omega)^{1/3}$, related to $d$ via
\begin{align}
d = \left( \frac{G (m_1+m_2)}{\Omega^2} \right)^{1/3}
 \equiv G (m_1+m_2) u^{-2}
\end{align}
we can rewrite the power as
\begin{equation} \label{eq:new_scalar_power}
P_{n=2, m=0} = \frac{\pi \beta_\mathrm{binary}^2}{2G} \left( \frac{r}{r_*} \right)^{3/2} \left( \frac{G(m_1+m_2)}{r} \right)^5 u^{-5} X^2(1-X)^2  J_{3/4}^2\left( \sqrt{3} r \Omega \right) \sin^2(\hat b) \; .
\end{equation}

The negative power in front of $u$ could seem worrisome as it diverges for small velocities, but recall that we are expanding in $d/r$ so $d$ cannot be taken to infinity without invalidating our expansion. Let us compare this expression with the GR quadrupole of a binary system in isolation $P_\mathrm{GR} \sim u^{10}/G$. Expanding the Bessel function for small arguments, we find
\begin{equation}
\frac{P_{n=2, m=0}}{P_\mathrm{GR}} \sim \beta_\mathrm{binary}^2 \left( \frac{r}{r_*} \right)^{3/2} \left( \frac{G(m_1+m_2)}{r} \right)^{7/2} u^{-21/2}
\end{equation}

With the orders-of-magnitude given in the introduction, one can evaluate this ratio to be
\begin{equation}
\frac{P_{n=2, m=0}}{P_\mathrm{GR}} \sim 10^{3} \beta_\mathrm{binary}^2 \left( \frac{r}{r_*} \right)^{3/2}
\end{equation}

Just as we proved in Section \ref{sec:dip_n0} that the Vainshtein screening is efficient for the motion of the barycentre of the system around the central mass, we see  that it is also  efficient for the motion of the binary system itself apart from a small enhancement factor of order $10^3$ ( recall that $(r/r_*)^{3/2} \sim 10^{-16}$ for a cosmological Galileon). In the next Section, we will compute the modification to the phase induced by this new power loss by assuming it is a small correction to the GR quadrupole.

%In particular for large enough $\beta_\mathrm{eff}\gtrsim 10^3$, the power emitted in the scalar radiation modes can be larger than the quadrupole radiation from GR. We will investigate this regime further below.

Finally, one can also compute the term in the dissipated power coming from the harmonics $m=\pm 1$ with the same strategy, giving a result
\begin{align}
 \begin{split}
P_{n=2, m=\pm 1} &= \frac{9 \beta_\mathrm{binary}^2}{512} \frac{(m_1+m_2)^2}{\mpl^2} \frac{d^4}{r^{7/2} r_*^{3/2}} X^2(1-X)^2 \sin^4(\hat b) \\
& \times \left((2 \Omega + \Omega_0)  J_{3/4}^2\left(\frac{ \sqrt{3}}{2} r (2 \Omega + \Omega_0) \right) + \vert 2 \Omega - \Omega_0 \vert  J_{3/4}^2\left(\frac{ \sqrt{3}}{2} r \vert 2 \Omega - \Omega_0 \vert \right) \right) \; .
\end{split}
\end{align}
This term involves a mixing between the two-body trajectory (parameterised by $\Omega$) and the revolution around the central black hole (parameterised by $\Omega_0$).  With the conditions $\Omega_0 \lesssim  \Omega$ and $d/r \lesssim 1$, this is a subleading correction compared to the $n=0$ term~\eqref{eq:dipole_n0term}.

\section{Inspirals and scalar correction to the phase} \label{sec:inspiral}
%and place ourselves in the situation where $r$ can be taken as constant (in other words, the binary inspiral induced by the scalar power is sufficiently quick so that we can ignore the motion around the central black hole)
Here we compute the scalar correction to the GW phase recorded in a detector such as LISA. Within our simple approximation scheme, there are two kind of GW signals : the one caused by the motion of the centre-of-mass around the supermassive BH (at a frequency $\Omega_0$), as well as the motion of the binary around its own centre-of-mass (at a frequency $\Omega$). We make the assumption that each of these two motions can be treated separately - the so-called "Hill's approximation" - and we compute the relevant scalar phase in the two following subsections.

\subsection{Extreme mass-ratio inspiral} \label{sec:EMRI}

In this Section, we will compute the modification to the GW phase (due to scalar wave emission) of the inspiral of the binary as a whole around the massive BH. Alternatively, our equations will be valid for any extreme-mass ratio inspiral of a light object around a massive BH provided once replaces the total mass $m_1+m_2$ by the mass of the small object. This will allow us to derive a lower bound on the effective coupling $\beta_\mathrm{binary}$ in order to be able to detect a Galileon correction to the GW phase.
% {\bf This corresponds to the case discussed in the previous section where the binary system falls more rapidly towards the central BH than onto itself. Notice that  this will allow us to derive a lower bound on the effective coupling $\beta$, i.e. the coupling of the two small BH to the scalar field,  to detect a Galileon correction to the GW phase.}

Since the scalar power loss is enhanced by the fact that it is a dipole, we will compute the phase assuming that it induces the leading deviation from GR (the scalar correction to the energy is 'usually' Vainshtein suppressed, i.e. it does not involve a dipolar enhancement, so we will neglect it).
The relevant dipolar power correction is given by equation \eqref{eq:dipole_n0term}. We still assume that the GR quadrupole dominates the power loss.

Using Kepler's third law, the Newtonian energy of the system is
\be
E = - \frac{(m_1+m_2) v^2}{2 } \; ,
\ee
where we recall that $v = (Gm_0 \Omega_0)^{1/3}$. The time evolution of the binary system is given by the balance equation
\begin{equation} \label{eq:balance}
\frac{dE}{dt}= -P_{GR}- P_\pi \; ,
\end{equation}
where $P_\pi$ is the scalar power loss for the $n=0$ dipole term in Eq. \eqref{eq:dipole_n0term} (with the Bessel function expanded for small arguments) and $P_\mathrm{GR}$ is the GR quadrupolar power loss given by
\begin{equation}
P_\mathrm{GR} = \frac{32}{5G} \left( \frac{m_1+m_2}{m_0} \right)^2 v^{10}
\end{equation}
Let us introduce the dimensionless constant $C_1$ by $P_\pi = C_1 P_\mathrm{GR} v^{-13/2}$, i.e.
\begin{equation}
C_1 = \frac{5 \times 3^{3/2} \pi \beta_\mathrm{binary}^2}{128 \Gamma(7/4)^2} \left( \frac{Gm_0}{r_*} \right)^{3/2} \; .
\end{equation}
The balance equation \eqref{eq:balance} then takes the form
\begin{equation}
\frac{dv}{dt} = \frac{32 (m_1+m_2)}{5 G m_0^2} v^9 \left( 1 + C_1 v^{-13/2} \right) \; .
\end{equation}
The number of observable GW cycles of the binary system in the detector is $2 \Phi(t)$ where
\begin{equation}
\Phi(t)= \int_{t_\mathrm{in}}^t \Omega_0 dt= \frac{1}{Gm_0} \int_{v_\mathrm{in}}^v dv v^3 \left(\frac{dt}{dv}\right) \; ,
\end{equation}
where $t_\mathrm{in}$ is the initial time  and $v_\mathrm{in}=v(t_\mathrm{in})$.

 We concentrate on initial conditions such that the two objects are separated by their redressed Vainshtein radii when our perturbative calculation is valid, i.e.
$r_{\rm in}\simeq ({\beta_0 m_0}/\beta_\mathrm{binary} m_{1,2})^{1/3}d$. As $r$ decreases, the binary system eventually disintegrate because of tidal effects at a final radius
$r_{\rm out}\simeq (m_0/m_{1,2})^{1/3} d$ implying that
\be
v_\mathrm{out} \sim (\beta_0/\beta_\mathrm{binary})^{1/6} v_\mathrm{in} \; .
\ee
The total accumulated phase during the inspiral is $\Phi = \Phi_\mathrm{GR} + \Delta \Phi$ where $\Phi_\mathrm{GR}$ is the usual GR phase,
\begin{equation}
\Phi_\mathrm{GR} = \frac{1}{32} \frac{m_0}{m_1+m_2} (v_\mathrm{in}^{-5}-v_\mathrm{out}^{-5})\simeq  \frac{1}{32} \frac{m_0}{m_1+m_2} v_\mathrm{in}^{-5} \sim 10^7 \; ,
\end{equation}
for $\Omega_0\simeq 10^{-3} \mathrm{Hz}$ as long as $\beta_0\gg \beta\mathrm{binary}$.
Now $\Delta \Phi$ is the correction due to the scalar field and reads
\begin{equation}
\Delta \Phi = \frac{10}{736}  \frac{m_0}{m_1+m_2} C_1 v_\mathrm{in}^{-23/2} \; .
\end{equation}
With the order-of-magnitudes given in the introduction and expanding in all the relevant parameters, we find
\begin{equation} \label{eq:delta_phi_EMRI}
\Delta \Phi \simeq 10^{-6} \beta_\mathrm{binary}^{2} \beta_0^{-1/2} \left( \frac{\Lambda}{10^{-12} \mathrm{eV}} \right)^{3/2} \; \left( \frac{m_1+m_2}{60 M_\odot} \right)^{-1} \; \left( \frac{m_0}{10^6 M_\odot} \right)^{-11/6} \; \left( \frac{\Omega_{0,\mathrm{in}}}{10^{-3} \mathrm{Hz}} \right)^{-23/6} \; ,
\end{equation}
where $\Omega_{0,\mathrm{in}}$ is the minimal frequency at which the GW signal is observed in the detector.

The precision achieved on the phase of GW observatories is at the level of $\Delta \Phi \sim 0.1$ \cite{Babak_2017}, so that an effective scalar charge $ \beta_\mathrm{binary}^{2} \beta_0^{-1/2} \gtrsim 10^6$ would induce observable modifications to the GW phase. Note that $\Delta \Phi$ is quite sensitive to the minimal frequency $\Omega_\mathrm{in}$ at which the signal is detected in the interferometer, so that a lower $\Omega_\mathrm{in}$ would greatly increase the detectability of such an event. Likewise, a value of $\Lambda \sim 10^{-8}$ eV would allow for a detection with an effective scalar charge of order one.

\subsection{Binary inspiral}

We now concentrate on the GW signal originating from the motion of the binary system around its centre-of-mass, entering a detector at frequency $\Omega$.
We make the assumption that the new formula we found for the enhanced scalar dissipated power in Eq. \eqref{eq:new_scalar_power} dominates the scalar loss of energy of the binary system, while the conservative energy is still dominated by GR, since the scalar energy is Vainshtein suppressed from Eq. \eqref{eq:energy}. We still assume that the GR quadrupole dominates the power loss and we will compute the modification to the phase of the GW as in the previous Section.

% This can be the case for moderately large values of the effective coupling
%$\beta_{\rm eff}$. More precisely and using Kepler's third law the energy of the binary system can be written as

The Newtonian energy of the binary system is
\be
E = - \frac{(m_1+m_2) \nu u^2}{2} \; ,
\ee
where $\nu=\frac{\mu}{m_1+m_2}$, $\mu=\frac{m_1m_2}{m_1+m_2}$ and we recall that $u = \left( G (m_1+m_2) \Omega \right)^{1/3}$. The balance equation takes the same form,
\be
\frac{dE}{dt}= -P_{GR}- P_\pi \; ,
\ee
where $P_\pi$ is the scalar power loss for the $m=0$ dipole term in Eq. \eqref{eq:new_scalar_power} and $P_\mathrm{GR}$ is the GR quadrupolar power loss given by
\begin{equation}
P_\mathrm{GR} = \frac{32}{5G} \nu^2 u^{10}
\end{equation}
 Let us introduce the dimensionless 'constant' $C_2$ by writing $P_\pi= C_2 \nu^2 u^{-1/2} / G$, i.e
\begin{equation}
C_2=\frac{3^{3/4} \pi \beta_\mathrm{binary}^2}{4 \sqrt{2} \Gamma(7/4)^2} \left( \frac{r}{r_*} \right)^{3/2} \left( \frac{G(m_1+m_2)}{r} \right)^{7/2}   \sin^2(\hat b)
\end{equation}
where once again we have expanded the Bessel function in eq. \eqref{eq:new_scalar_power} for small arguments. Then the time evolution of $u$ is given by
\begin{equation} \label{eq:dudt}
\frac{d u}{d t} = \frac{\nu}{G(m_1+m_2)} \left( \frac{32}{5} u^9 + C_2 u^{-1/2} \right)
\end{equation}

In the regime where the first term coming from the GR quadrupole dominates, we have
\begin{equation} \label{eq:ut}
u(t)= u_\mathrm{in}\left( 1-  \frac{(t-t_\mathrm{in})}{t_1} \right)^{-1/8} \; ,
\end{equation}
where
\be
t_1=\frac{5}{256} \frac{G(m_1+m_2)}{\nu u_\mathrm{in}^8}=\frac{5}{256}\frac{d_\mathrm{in}^4}{G^3 (m_1+m_2)^3 \nu} \; ,
\ee
is the time of collapse from the initial time $t_\mathrm{in}$ where $u_\mathrm{in}=u(t_\mathrm{in})$, $d_\mathrm{in}=d(t_\mathrm{in})$, which we choose to be the time when our perturbative calculation becomes valid according to Eq. \eqref{eq:condition_d}.

However, the binary system  always falls into the central BH before merging. This is because the time of collapse into the central BH is
\begin{equation}
t_0 = \frac{5}{256} \frac{Gm_0^2}{(m_1+m_2) v_\mathrm{in}^8} \; ,
\end{equation}
as the reduced mass is $m_1+m_2$ and the mass of the central object is $m_0$, and we recall that we have introduced the variable $v=(G m_0 \Omega_0)^{1/3}$. Since at the initial time where the perturbative calculation begins to be valid one has $\Omega^\mathrm{in} = (\beta_0 / \beta_\mathrm{binary})^{1/2} \Omega_0^\mathrm{in}$ (see Eq. \eqref{eq:bound_omega}), the ratio of the collapse times is
\begin{equation}
\frac{t_0}{t_1} = \left( \frac{m_1 + m_2}{m_0} \right)^{2/3} \left( \frac{\beta_0}{\beta_\mathrm{binary}} \right)^{4/3}  \; .
\end{equation}
With the moderately large ratio of couplings we choose to study in this article (see Eq.~\eqref{eq:r_omega}), this factor is less than one. Thus, one would not observe the merging of the binary system itself as  it will be destroyed by tidal disruption before being swallowed by the supermassive BH. More precisely,
using the  simplifying assumption that $r$ decreases according to the quadrupole formula, i.e.
\begin{equation}
r = r_\mathrm{in} \left( 1 - \frac{t-t_\mathrm{in}}{t_0} \right)^{1/4},
\end{equation}
we see that tidal effects become relevant and make the binary system explode after an interval of time
\be
\frac{\Delta t}{t_0}\simeq 1- \left(\frac{\beta}{\beta_0}\right)^{4/3}
\ee
which is of order $t_0$.
Before being destroyed by tidal effects, the binary system produces GW. The number of observable GW cycles of the binary system in the detector is $2 \Phi(t)$ where
\begin{equation}
\Phi(t)= \int_{t_\mathrm{in}}^t \Omega dt= \frac{1}{G(m_1+m_2)} \int_{u_\mathrm{in}}^u du u^3 \left(\frac{dt}{du}\right) \; .
\end{equation}
Let us compute first the number of cycles assuming that the power loss is purely GR. One finds
\begin{equation}
\Phi = \frac{1}{32 \nu} \left( u_\mathrm{in}^{-5} - u_\mathrm{out}^{-5} \right) \; ,
\end{equation}
where $u_\mathrm{out} = u(t_0)$ is the value of $u$ when the binary system explodes due to the tidal effects. Expanding Eq. \eqref{eq:ut} for $t_0 / t_1 \ll 1$, one finds
\begin{equation}
\Phi \simeq \frac{5}{256} u_\mathrm{in}^{-5} \frac{t_0}{t_1} \gtrsim 10^5 \; ,
\end{equation}
which shows that the signal would be observable during a fairly large amount of cycles. Let us now compute the correction to the phase due to the scalar power loss. Expanding equation \eqref{eq:dudt} for small $C_2$, it is
\begin{equation}
\Delta \Phi = \frac{25}{1024 \nu} \int \mathrm{d} u \; C_2 u^{-31/2}
\end{equation}
In the integral, $C_2$ depends on time (so it depends on $u$) \textit{via} its dependence on $r$.
We can relate $t-t_\mathrm{in}$ to $u$ using Eq. \eqref{eq:ut} expanded for $t/t_1 \ll 1$. Since $C_2 \propto r^{-2}$, this gives
\begin{equation}
\Delta \Phi = \frac{25}{1024 \nu} C_\mathrm{2, \; in} u_\mathrm{in}^{-29/2} \int_1^{1+\frac{t_0}{8t_1}} \mathrm{d} x \; x^{-31/2} \left(1 - 8 \frac{t_1}{t_0} \left(x - 1 \right)  \right)^{-1/2} \; ,
\end{equation}
where we have used the change of variable $x = u/u_\mathrm{in}$. Expanding once again for $t_0/t_1 \ll 1$, we finally find
\begin{equation}
\Delta \Phi = \frac{25}{4096 \nu} C_\mathrm{2, \; in} u_\mathrm{in}^{-29/2} \frac{t_0}{t_1} \; .
\end{equation}
With the order-of-magnitudes given in the introduction and expanding in all the relevant parameters, we find
\begin{equation}
\Delta \Phi \simeq 10^{-7} \beta_\mathrm{binary}^{4/3} \beta_0^{1/6}   \left( \frac{\Lambda}{10^{-12} \mathrm{eV}} \right)^{3/2} \; \left( \frac{m_1+m_2}{60 M_\odot} \right)^{-2/3} \; \left( \frac{m_0}{10^6 M_\odot} \right)^{-11/6} \; \left( \frac{\Omega_\mathrm{in}}{10^{-3} \mathrm{Hz}} \right)^{-7/2}
\end{equation}
This result is similar to the one that we obtained above \eqref{eq:delta_phi_EMRI}, with slightly different powers of the parameters. Again for an effective scalar charge $\beta_\mathrm{binary}^{4/3} \beta_0^{1/6} \gtrsim 10^7$ we could expect to detect the inspiralling phase of the two BHs falling towards each other.

\section{Conclusions}
\label{SecCon}
Galileons  give rise  to scalar interactions that are embedded  in prominent effective field theories of gravity like massive gravity, generalised Proca and Horndeski. Looking for experimental  evidence of Galileon interactions is a crucial step in testing these theories. Cosmological and astrophysical phenomena provide important testbeds in that respect. Given the fast and promising developments in  GW physics in the last few  years, we focused here on a situation where a Galileon interaction could lead to modifications of  GW signals potentially detectable by laser interferometers. To be specific, we chose to concentrate on the cubic Galileon model, for simplicity's sake, as it captures the relevant features of higher derivative self-interactions. The three-body system we have considered comprises a supermassive BH in the centre of a galaxy and two BHs forming a binary system in its vicinity.

We focused on a regime where perturbative calculations are applicable and computed the cubic Galileon corrections to the orbital parameters of the binary system. To do so, we considered the small perturbations due to the binary system in the  background of the central BH. The two small BHs rotate with a frequency $\Omega$ around their common barycentre, which itself orbits with a frequency $\Omega_0$ around the supermassive BH. We expanded the action to cubic order in perturbations due to the binary system. The quadratic terms enabled us to determine the Green's function and hence the Galileon propagator in the background of the central BH. With the help of the Green's function we were able to compute explicitly the two-body energy and the power dissipated in scalar radiation. The cubic self-interactions due to the binary system are negligible when the distance between the BHs is larger than their redressed Vainshtein radii in the background of the central BH. These results require the existence of a non-trivial conformal coupling of the scalar field to matter which is taken to have a dynamical origin, i.e. it appears when the background field is explicitly time-dependent.

Starting with the static limit, we  found that the two-body scalar energy of the binary system is suppressed by the Vainshtein factor corresponding to the distance to the supermassive BH. In the case of circular orbits, we showed that the effect of the scalar energy translates into a renormalisation of  Newton's constant. As expected, the two-body scalar energy gives a correction proportional to the effective matter coupling of the Galileon $\beta_\mathrm{eff}^2$ and is screened by the Vainshtein factor $(r/r_*)^{3/2}$ due to the central BH, where $r$ is the distance of the binary system to the central BH. Then we computed the dissipated power in the presence of the cubic Galileon. Due to the absence of non-minimal couplings to matter, the tensor radiation of GW waves follows the standard quadrupole formula. Crucial new effects appear  in the scalar radiation. We found that the scalar  monopole  only depends on $(d\Omega)^2$ and not on $(d/r)^2$. Concerning the dipole moment, we computed both the dipolar emission of the centre-of-mass due to its motion around the central BH and the dipolar emission from the motion of the small masses themselves around their common  centre-of-mass. In the former, the dipole power is enhanced compared to the GR case by two powers of the velocity. Unfortunately it is still suppressed by the Vainshtein factor $1/(\Omega_0 r_*)^{3/2}$ implying that the overall scalar emission is suppressed. In the latter, we  found a velocity enhancement due to the dipolar power leading to a small reduction of the Vainshtein efficiency.
% On the contrary, the scalar emission of scalars by the binary system itself can dominate over the GR emission rate for large enough conformal couplings. In this case, the effects of the scalar field can be twofold. If the coupling is not too large, the binary system actually merges with the central BH before collapsing onto itself. In this scalar dominated phase, the number of observable cycles can be large. If the coupling is even larger, say larger than $10^5$, the binary system is still absorbed by the central BH and only the beginning of the inspiralling phase is dominated by the scalar field. Soon enough, the two BH's behave  like in GR. Hence we have found that, for certain values of the conformal couplings, and  for certain triple systems where the distance between the two BHs in the binary system is large enough to avoid the Vainshtein screening, the Galileon scalar field radiation could be the dominant effect in the dynamics of binary systems.

Finally, we have derived the Galileon corrections to the phase of gravitational waves emitted from the system. Since the number of cycles can be very large, the relative precision of GW observatories could be excellent allowing for detection of  tiny modifications from GR. In this respect, we have found that for large enough conformal couplings a modification to the GW phase due to the Galileon can be induced. Our results can alternatively be seen as constraints on the strong coupling scale $\Lambda$ present in all Galileon interactions.

 Of course, the configuration studied in this article is very special. For BHs of masses around 30 $M_\odot$ and for frequency in the LISA band, this requires that the two BHs in the binary system should be closer than 0.01 AU and that they should  be around   0.5 AU away from the central BH ; furthermore, a more realistic treatment would necessitate to take into account the higher order post-Newtonian corrections, the eccentricity of the orbit, the spins of the BH, and the accretion disk of the supermassive BH \cite{McKernan:2019beu}. Part of our results are also valid for the EMRI of a single BH around a supermassive one (see Section \ref{sec:EMRI}), which are among the main targets of LISA.
For such configurations, the presence of a Galileon field could be hoped to be detected.

\acknowledgments
We would like to thank Federico Piazza for discussions. LH is supported by funding from the European Research Council (ERC) under the European Unions Horizon 2020 research and innovation programme grant agreement No 801781 and by the Swiss National Science Foundation grant 179740. This work is
supported in part by the EU Horizon 2020 research and innovation
programme under the Marie-Sklodowska grant No. 690575. This article is
based upon work related to the COST Action CA15117 (CANTATA) supported
by COST (European Cooperation in Science and Technology).

\bibliographystyle{JHEP}
\bibliography{Gal_3Body}{}

\providecommand{\href}[2]{#2}\begingroup\raggedright\begin{thebibliography}{10}

\bibitem{BeltranJimenez:2019tjy}
J.~B. Jim{\'e}nez, L.~Heisenberg and T.~S. Koivisto, \emph{{The Geometrical
  Trinity of Gravity}},
  \href{https://doi.org/10.3390/universe5070173}{\emph{Universe} {\bfseries 5}
  (2019) 173} [\href{https://arxiv.org/abs/1903.06830}{{\ttfamily
  1903.06830}}].

\bibitem{Akiyama:2019cqa}
{\scshape Event Horizon Telescope} collaboration, K.~Akiyama et~al.,
  \emph{{First M87 Event Horizon Telescope Results. I. The Shadow of the
  Supermassive Black Hole}},
  \href{https://doi.org/10.3847/2041-8213/ab0ec7}{\emph{Astrophys. J.}
  {\bfseries 875} (2019) L1}
  [\href{https://arxiv.org/abs/1906.11238}{{\ttfamily 1906.11238}}].

\bibitem{Wex:2014nva}
N.~Wex, \emph{{Testing Relativistic Gravity with Radio Pulsars}},
  \href{https://arxiv.org/abs/1402.5594}{{\ttfamily 1402.5594}}.

\bibitem{Abbott:2016blz}
{\scshape LIGO Scientific, Virgo} collaboration, B.~P. Abbott et~al.,
  \emph{{Observation of Gravitational Waves from a Binary Black Hole Merger}},
  \href{https://doi.org/10.1103/PhysRevLett.116.061102}{\emph{Phys. Rev. Lett.}
  {\bfseries 116} (2016) 061102}
  [\href{https://arxiv.org/abs/1602.03837}{{\ttfamily 1602.03837}}].

\bibitem{TheLIGOScientific:2017qsa}
{\scshape LIGO Scientific, Virgo} collaboration, B.~P. Abbott et~al.,
  \emph{{GW170817: Observation of Gravitational Waves from a Binary Neutron
  Star Inspiral}},
  \href{https://doi.org/10.1103/PhysRevLett.119.161101}{\emph{Phys. Rev. Lett.}
  {\bfseries 119} (2017) 161101}
  [\href{https://arxiv.org/abs/1710.05832}{{\ttfamily 1710.05832}}].

\bibitem{Brax:2015dma}
P.~Brax, C.~Burrage and A.-C. Davis, \emph{{The Speed of Galileon Gravity}},
  \href{https://doi.org/10.1088/1475-7516/2016/03/004}{\emph{JCAP} {\bfseries
  1603} (2016) 004} [\href{https://arxiv.org/abs/1510.03701}{{\ttfamily
  1510.03701}}].

\bibitem{Lombriser:2016yzn}
L.~Lombriser and N.~A. Lima, \emph{{Challenges to Self-Acceleration in Modified
  Gravity from Gravitational Waves and Large-Scale Structure}},
  \href{https://doi.org/10.1016/j.physletb.2016.12.048}{\emph{Phys. Lett.}
  {\bfseries B765} (2017) 382}
  [\href{https://arxiv.org/abs/1602.07670}{{\ttfamily 1602.07670}}].

\bibitem{Ezquiaga:2017ekz}
J.~M. Ezquiaga and M.~Zumalac{\'a}rregui, \emph{{Dark Energy After GW170817:
  Dead Ends and the Road Ahead}},
  \href{https://doi.org/10.1103/PhysRevLett.119.251304}{\emph{Phys. Rev. Lett.}
  {\bfseries 119} (2017) 251304}
  [\href{https://arxiv.org/abs/1710.05901}{{\ttfamily 1710.05901}}].

\bibitem{Creminelli:2017sry}
P.~Creminelli and F.~Vernizzi, \emph{{Dark Energy after GW170817 and
  GRB170817A}},
  \href{https://doi.org/10.1103/PhysRevLett.119.251302}{\emph{Phys. Rev. Lett.}
  {\bfseries 119} (2017) 251302}
  [\href{https://arxiv.org/abs/1710.05877}{{\ttfamily 1710.05877}}].

\bibitem{Sakstein:2017xjx}
J.~Sakstein and B.~Jain, \emph{{Implications of the Neutron Star Merger
  GW170817 for Cosmological Scalar-Tensor Theories}},
  \href{https://doi.org/10.1103/PhysRevLett.119.251303}{\emph{Phys. Rev. Lett.}
  {\bfseries 119} (2017) 251303}
  [\href{https://arxiv.org/abs/1710.05893}{{\ttfamily 1710.05893}}].

\bibitem{Baker:2017hug}
T.~Baker, E.~Bellini, P.~G. Ferreira, M.~Lagos, J.~Noller and I.~Sawicki,
  \emph{{Strong constraints on cosmological gravity from GW170817 and GRB
  170817A}}, \href{https://doi.org/10.1103/PhysRevLett.119.251301}{\emph{Phys.
  Rev. Lett.} {\bfseries 119} (2017) 251301}
  [\href{https://arxiv.org/abs/1710.06394}{{\ttfamily 1710.06394}}].

\bibitem{Heisenberg:2018vsk}
L.~Heisenberg, \emph{{A systematic approach to generalisations of General
  Relativity and their cosmological implications}},
  \href{https://doi.org/10.1016/j.physrep.2018.11.006}{\emph{Phys. Rept.}
  {\bfseries 796} (2019) 1} [\href{https://arxiv.org/abs/1807.01725}{{\ttfamily
  1807.01725}}].

\bibitem{Ezquiaga:2018btd}
J.~M. Ezquiaga and M.~Zumalac{\'a}rregui, \emph{{Dark Energy in light of
  Multi-Messenger Gravitational-Wave astronomy}},
  \href{https://doi.org/10.3389/fspas.2018.00044}{\emph{Front. Astron. Space
  Sci.} {\bfseries 5} (2018) 44}
  [\href{https://arxiv.org/abs/1807.09241}{{\ttfamily 1807.09241}}].

\bibitem{Kase:2018aps}
R.~Kase and S.~Tsujikawa, \emph{{Dark energy in Horndeski theories after
  GW170817: A review}},
  \href{https://doi.org/10.1142/S0218271819420057}{\emph{Int. J. Mod. Phys.}
  {\bfseries D28} (2019) 1942005}
  [\href{https://arxiv.org/abs/1809.08735}{{\ttfamily 1809.08735}}].

\bibitem{Horndeski:1974wa}
G.~W. Horndeski, \emph{{Second-order scalar-tensor field equations in a
  four-dimensional space}},
  \href{https://doi.org/10.1007/BF01807638}{\emph{Int. J. Theor. Phys.}
  {\bfseries 10} (1974) 363}.

\bibitem{Heisenberg:2014rta}
L.~Heisenberg, \emph{{Generalization of the Proca Action}},
  \href{https://doi.org/10.1088/1475-7516/2014/05/015}{\emph{JCAP} {\bfseries
  1405} (2014) 015} [\href{https://arxiv.org/abs/1402.7026}{{\ttfamily
  1402.7026}}].

\bibitem{Jimenez:2016isa}
J.~Beltran~Jimenez and L.~Heisenberg, \emph{{Derivative self-interactions for a
  massive vector field}},
  \href{https://doi.org/10.1016/j.physletb.2016.04.017}{\emph{Phys. Lett.}
  {\bfseries B757} (2016) 405}
  [\href{https://arxiv.org/abs/1602.03410}{{\ttfamily 1602.03410}}].

\bibitem{deRham:2010kj}
C.~de~Rham, G.~Gabadadze and A.~J. Tolley, \emph{{Resummation of Massive
  Gravity}}, \href{https://doi.org/10.1103/PhysRevLett.106.231101}{\emph{Phys.
  Rev. Lett.} {\bfseries 106} (2011) 231101}
  [\href{https://arxiv.org/abs/1011.1232}{{\ttfamily 1011.1232}}].

\bibitem{Nicolis:2008in}
A.~Nicolis, R.~Rattazzi and E.~Trincherini, \emph{{The Galileon as a local
  modification of gravity}},
  \href{https://doi.org/10.1103/PhysRevD.79.064036}{\emph{Phys. Rev.}
  {\bfseries D79} (2009) 064036}
  [\href{https://arxiv.org/abs/0811.2197}{{\ttfamily 0811.2197}}].

\bibitem{Nicolis:2004qq}
A.~Nicolis and R.~Rattazzi, \emph{{Classical and quantum consistency of the DGP
  model}}, \href{https://doi.org/10.1088/1126-6708/2004/06/059}{\emph{JHEP}
  {\bfseries 06} (2004) 059}
  [\href{https://arxiv.org/abs/hep-th/0404159}{{\ttfamily hep-th/0404159}}].

\bibitem{Hinterbichler:2010xn}
K.~Hinterbichler, M.~Trodden and D.~Wesley, \emph{{Multi-field galileons and
  higher co-dimension branes}},
  \href{https://doi.org/10.1103/PhysRevD.82.124018}{\emph{Phys. Rev.}
  {\bfseries D82} (2010) 124018}
  [\href{https://arxiv.org/abs/1008.1305}{{\ttfamily 1008.1305}}].

\bibitem{dePaulaNetto:2012hm}
T.~de~Paula~Netto and I.~L. Shapiro, \emph{{One-loop divergences in the
  Galileon model}},
  \href{https://doi.org/10.1016/j.physletb.2012.08.056}{\emph{Phys. Lett.}
  {\bfseries B716} (2012) 454}
  [\href{https://arxiv.org/abs/1207.0534}{{\ttfamily 1207.0534}}].

\bibitem{Heisenberg:2014raa}
L.~Heisenberg, \emph{{Quantum Corrections in Galileons from Matter Loops}},
  \href{https://doi.org/10.1103/PhysRevD.90.064005}{\emph{Phys. Rev.}
  {\bfseries D90} (2014) 064005}
  [\href{https://arxiv.org/abs/1408.0267}{{\ttfamily 1408.0267}}].

\bibitem{Heisenberg:2019udf}
L.~Heisenberg and C.~F. Steinwachs, \emph{{One-loop renormalization in Galileon
  effective field theory}},
  \href{https://doi.org/10.1088/1475-7516/2020/01/014}{\emph{JCAP} {\bfseries
  2001} (2020) 014} [\href{https://arxiv.org/abs/1909.04662}{{\ttfamily
  1909.04662}}].

\bibitem{Heisenberg:2019wjv}
L.~Heisenberg and C.~F. Steinwachs, \emph{{Geometrized quantum Galileons}},
  \href{https://arxiv.org/abs/1909.07111}{{\ttfamily 1909.07111}}.

\bibitem{Vainshtein:1972sx}
A.~I. Vainshtein, \emph{{To the problem of nonvanishing gravitation mass}},
  \href{https://doi.org/10.1016/0370-2693(72)90147-5}{\emph{Phys. Lett.}
  {\bfseries 39B} (1972) 393}.

\bibitem{Chow:2009fm}
N.~Chow and J.~Khoury, \emph{{Galileon Cosmology}},
  \href{https://doi.org/10.1103/PhysRevD.80.024037}{\emph{Phys. Rev.}
  {\bfseries D80} (2009) 024037}
  [\href{https://arxiv.org/abs/0905.1325}{{\ttfamily 0905.1325}}].

\bibitem{DeFelice:2010nf}
A.~De~Felice and S.~Tsujikawa, \emph{{Generalized Galileon cosmology}},
  \href{https://doi.org/10.1103/PhysRevD.84.124029}{\emph{Phys. Rev.}
  {\bfseries D84} (2011) 124029}
  [\href{https://arxiv.org/abs/1008.4236}{{\ttfamily 1008.4236}}].

\bibitem{deRham:2011by}
C.~de~Rham and L.~Heisenberg, \emph{{Cosmology of the Galileon from Massive
  Gravity}}, \href{https://doi.org/10.1103/PhysRevD.84.043503}{\emph{Phys.
  Rev.} {\bfseries D84} (2011) 043503}
  [\href{https://arxiv.org/abs/1106.3312}{{\ttfamily 1106.3312}}].

\bibitem{deRham:2010tw}
C.~de~Rham, G.~Gabadadze, L.~Heisenberg and D.~Pirtskhalava, \emph{{Cosmic
  Acceleration and the Helicity-0 Graviton}},
  \href{https://doi.org/10.1103/PhysRevD.83.103516}{\emph{Phys. Rev.}
  {\bfseries D83} (2011) 103516}
  [\href{https://arxiv.org/abs/1010.1780}{{\ttfamily 1010.1780}}].

\bibitem{Creminelli:2010ba}
P.~Creminelli, A.~Nicolis and E.~Trincherini, \emph{{Galilean Genesis: An
  Alternative to inflation}},
  \href{https://doi.org/10.1088/1475-7516/2010/11/021}{\emph{JCAP} {\bfseries
  1011} (2010) 021} [\href{https://arxiv.org/abs/1007.0027}{{\ttfamily
  1007.0027}}].

\bibitem{LevasseurPerreault:2011mw}
L.~Perreault~Levasseur, R.~Brandenberger and A.-C. Davis, \emph{{Defrosting in
  an Emergent Galileon Cosmology}},
  \href{https://doi.org/10.1103/PhysRevD.84.103512}{\emph{Phys. Rev.}
  {\bfseries D84} (2011) 103512}
  [\href{https://arxiv.org/abs/1105.5649}{{\ttfamily 1105.5649}}].

\bibitem{Kobayashi:2010cm}
T.~Kobayashi, M.~Yamaguchi and J.~Yokoyama, \emph{{G-inflation: Inflation
  driven by the Galileon field}},
  \href{https://doi.org/10.1103/PhysRevLett.105.231302}{\emph{Phys. Rev. Lett.}
  {\bfseries 105} (2010) 231302}
  [\href{https://arxiv.org/abs/1008.0603}{{\ttfamily 1008.0603}}].

\bibitem{Burrage:2010cu}
C.~Burrage, C.~de~Rham, D.~Seery and A.~J. Tolley, \emph{{Galileon inflation}},
  \href{https://doi.org/10.1088/1475-7516/2011/01/014}{\emph{JCAP} {\bfseries
  1101} (2011) 014} [\href{https://arxiv.org/abs/1009.2497}{{\ttfamily
  1009.2497}}].

\bibitem{Brax:2011sv}
P.~Brax, C.~Burrage and A.-C. Davis, \emph{{Laboratory Tests of the Galileon}},
  \href{https://doi.org/10.1088/1475-7516/2011/09/020}{\emph{JCAP} {\bfseries
  1109} (2011) 020} [\href{https://arxiv.org/abs/1106.1573}{{\ttfamily
  1106.1573}}].

\bibitem{Babichev:2016fbg}
E.~Babichev, C.~Charmousis, A.~Lehbel and T.~Moskalets, \emph{{Black holes in a
  cubic Galileon universe}},
  \href{https://doi.org/10.1088/1475-7516/2016/09/011}{\emph{JCAP} {\bfseries
  1609} (2016) 011} [\href{https://arxiv.org/abs/1605.07438}{{\ttfamily
  1605.07438}}].

\bibitem{Babichev:2015rva}
E.~Babichev, C.~Charmousis and M.~Hassaine, \emph{{Charged Galileon black
  holes}}, \href{https://doi.org/10.1088/1475-7516/2015/05/031}{\emph{JCAP}
  {\bfseries 1505} (2015) 031}
  [\href{https://arxiv.org/abs/1503.02545}{{\ttfamily 1503.02545}}].

\bibitem{Wyman:2011mp}
M.~Wyman, \emph{{Galilean-invariant scalar fields can strengthen gravitational
  lensing}}, \href{https://doi.org/10.1103/PhysRevLett.106.201102}{\emph{Phys.
  Rev. Lett.} {\bfseries 106} (2011) 201102}
  [\href{https://arxiv.org/abs/1101.1295}{{\ttfamily 1101.1295}}].

\bibitem{Goon:2010xh}
G.~L. Goon, K.~Hinterbichler and M.~Trodden, \emph{{Stability and
  superluminality of spherical DBI galileon solutions}},
  \href{https://doi.org/10.1103/PhysRevD.83.085015}{\emph{Phys. Rev.}
  {\bfseries D83} (2011) 085015}
  [\href{https://arxiv.org/abs/1008.4580}{{\ttfamily 1008.4580}}].

\bibitem{deFromont:2013iwa}
P.~de~Fromont, C.~de~Rham, L.~Heisenberg and A.~Matas, \emph{{Superluminality
  in the Bi- and Multi- Galileon}},
  \href{https://doi.org/10.1007/JHEP07(2013)067}{\emph{JHEP} {\bfseries 07}
  (2013) 067} [\href{https://arxiv.org/abs/1303.0274}{{\ttfamily 1303.0274}}].

\bibitem{Seoane:2013qna}
{\scshape eLISA} collaboration, P.~A. Seoane et~al., \emph{{The Gravitational
  Universe}},  \href{https://arxiv.org/abs/1305.5720}{{\ttfamily 1305.5720}}.

\bibitem{Hawking1972}
S.~W. Hawking, \emph{Black holes in the brans-dicke},
  \href{https://doi.org/10.1007/BF01877518}{\emph{Communications in
  Mathematical Physics} {\bfseries 25} (1972) 167}.

\bibitem{Hui_2013}
L.~Hui and A.~Nicolis, \emph{No-hair theorem for the galileon},
  \href{https://doi.org/10.1103/physrevlett.110.241104}{\emph{Physical Review
  Letters} {\bfseries 110} (2013) }.

\bibitem{Jacobson:1999vr}
T.~Jacobson, \emph{{Primordial black hole evolution in tensor scalar
  cosmology}}, \href{https://doi.org/10.1103/PhysRevLett.83.2699}{\emph{Phys.
  Rev. Lett.} {\bfseries 83} (1999) 2699}
  [\href{https://arxiv.org/abs/astro-ph/9905303}{{\ttfamily
  astro-ph/9905303}}].

\bibitem{Horbatsch:2011ye}
M.~W. Horbatsch and C.~P. Burgess, \emph{{Cosmic Black-Hole Hair Growth and
  Quasar OJ287}},
  \href{https://doi.org/10.1088/1475-7516/2012/05/010}{\emph{JCAP} {\bfseries
  1205} (2012) 010} [\href{https://arxiv.org/abs/1111.4009}{{\ttfamily
  1111.4009}}].

\bibitem{deRham:2012fg}
C.~de~Rham, A.~Matas and A.~J. Tolley, \emph{{Galileon Radiation from Binary
  Systems}}, \href{https://doi.org/10.1103/PhysRevD.87.064024}{\emph{Phys.
  Rev.} {\bfseries D87} (2013) 064024}
  [\href{https://arxiv.org/abs/1212.5212}{{\ttfamily 1212.5212}}].

\bibitem{Dar:2018dra}
F.~Dar, C.~De~Rham, J.~T. Deskins, J.~T. Giblin and A.~J. Tolley, \emph{{Scalar
  Gravitational Radiation from Binaries: Vainshtein Mechanism in Time-dependent
  Systems}}, \href{https://doi.org/10.1088/1361-6382/aaf5e8}{\emph{Class.
  Quant. Grav.} {\bfseries 36} (2019) 025008}
  [\href{https://arxiv.org/abs/1808.02165}{{\ttfamily 1808.02165}}].

\bibitem{deRham:2012fw}
C.~de~Rham, A.~J. Tolley and D.~H. Wesley, \emph{{Vainshtein Mechanism in
  Binary Pulsars}},
  \href{https://doi.org/10.1103/PhysRevD.87.044025}{\emph{Phys. Rev.}
  {\bfseries D87} (2013) 044025}
  [\href{https://arxiv.org/abs/1208.0580}{{\ttfamily 1208.0580}}].

\bibitem{Eardley1975ApJ}
D.~M. {Eardley}, \emph{{Observable effects of a scalar gravitational field in a
  binary pulsar}}, \href{https://doi.org/10.1086/181744}{\emph{Astrophysical
  Journal} {\bfseries 196} (1975) L59}.

\bibitem{Kuntz:2019zef}
A.~Kuntz, F.~Piazza and F.~Vernizzi, \emph{{Effective field theory for
  gravitational radiation in scalar-tensor gravity}},
  \href{https://doi.org/10.1088/1475-7516/2019/05/052}{\emph{JCAP} {\bfseries
  1905} (2019) 052} [\href{https://arxiv.org/abs/1902.04941}{{\ttfamily
  1902.04941}}].

\bibitem{Randall:2019sab}
L.~Randall and Z.-Z. Xianyu, \emph{{Observing Eccentricity Oscillations of
  Binary Black Holes in LISA}},
  \href{https://arxiv.org/abs/1902.08604}{{\ttfamily 1902.08604}}.

\bibitem{Randall:2017jop}
L.~Randall and Z.-Z. Xianyu, \emph{{Induced Ellipticity for Inspiraling Binary
  Systems}}, \href{https://doi.org/10.3847/1538-4357/aaa1a2}{\emph{Astrophys.
  J.} {\bfseries 853} (2018) 93}
  [\href{https://arxiv.org/abs/1708.08569}{{\ttfamily 1708.08569}}].

\bibitem{Randall:2018nud}
L.~Randall and Z.-Z. Xianyu, \emph{{An Analytical Portrait of Binary Mergers in
  Hierarchical Triple Systems}},
  \href{https://doi.org/10.3847/1538-4357/aad7fe}{\emph{Astrophys. J.}
  {\bfseries 864} (2018) 134}
  [\href{https://arxiv.org/abs/1802.05718}{{\ttfamily 1802.05718}}].

\bibitem{Babichev:2013cya}
E.~Babichev and C.~Charmousis, \emph{{Dressing a black hole with a
  time-dependent Galileon}},
  \href{https://doi.org/10.1007/JHEP08(2014)106}{\emph{JHEP} {\bfseries 08}
  (2014) 106} [\href{https://arxiv.org/abs/1312.3204}{{\ttfamily 1312.3204}}].

\bibitem{Babichev_2013}
E.~Babichev and G.~Esposito-Far{\`e}se, \emph{Time-dependent spherically
  symmetric covariant galileons},
  \href{https://doi.org/10.1103/physrevd.87.044032}{\emph{Physical Review D}
  {\bfseries 87} (2013) }.

\bibitem{Abramowicz_2013}
M.~A. Abramowicz and P.~C. Fragile, \emph{Foundations of black hole accretion
  disk theory}, \href{https://doi.org/10.12942/lrr-2013-1}{\emph{Living Reviews
  in Relativity} {\bfseries 16} (2013) }.

\bibitem{WAGONER2008828}
R.~V. Wagoner, \emph{Relativistic and newtonian diskoseismology},
  \href{https://doi.org/https://doi.org/10.1016/j.newar.2008.03.012}{\emph{New
  Astronomy Reviews} {\bfseries 51} (2008) 828 }.

\bibitem{Kuntz:2019plo}
A.~Kuntz, \emph{{Two-body potential of Vainshtein screened theories}},
  \href{https://doi.org/10.1103/PhysRevD.100.024024}{\emph{Phys. Rev.}
  {\bfseries D100} (2019) 024024}
  [\href{https://arxiv.org/abs/1905.07340}{{\ttfamily 1905.07340}}].

\bibitem{Andrews_2013}
M.~Andrews, Y.-Z. Chu and M.~Trodden, \emph{Galileon forces in the solar
  system}, \href{https://doi.org/10.1103/physrevd.88.084028}{\emph{Physical
  Review D} {\bfseries 88} (2013) }.

\bibitem{Chu:2012kz}
Y.-Z. Chu and M.~Trodden, \emph{{Retarded Green's function of a Vainshtein
  system and Galileon waves}},
  \href{https://doi.org/10.1103/PhysRevD.87.024011}{\emph{Phys. Rev.}
  {\bfseries D87} (2013) 024011}
  [\href{https://arxiv.org/abs/1210.6651}{{\ttfamily 1210.6651}}].

\bibitem{Babak_2017}
S.~Babak, J.~Gair, A.~Sesana, E.~Barausse, C.~F. Sopuerta, C.~P. Berry et~al.,
  \emph{Science with the space-based interferometer lisa. v. extreme mass-ratio
  inspirals}, \href{https://doi.org/10.1103/physrevd.95.103012}{\emph{Physical
  Review D} {\bfseries 95} (2017) }.

\bibitem{McKernan:2019beu}
B.~McKernan, K.~E.~S. Ford, R.~O'Shaughnessy and D.~Wysocki, \emph{{Monte-Carlo
  simulations of black hole mergers in AGN disks: Low $\chi_{\rm eff}$ mergers
  and predictions for LIGO}},
  \href{https://arxiv.org/abs/1907.04356}{{\ttfamily 1907.04356}}.

\end{thebibliography}\endgroup
\end{document}